\documentclass[screen, sigconf]{acmart}

\AtBeginDocument{%
  }

\setcopyright{acmlicensed}
\copyrightyear{2026}
\acmYear{2026}
\setcopyright{cc}
\setcctype{by}
\acmConference[CHI '26]{Proceedings of the 2026 CHI Conference on Human Factors in Computing Systems}{April 13--17, 2026}{Barcelona, Spain}
\acmBooktitle{Proceedings of the 2026 CHI Conference on Human Factors in Computing Systems (CHI '26), April 13--17, 2026, Barcelona, Spain}
\acmPrice{}
\acmDOI{10.1145/3772318.3790292}
\acmISBN{979-8-4007-2278-3/2026/04}

\usepackage{subcaption}
\usepackage{multirow}
\usepackage{makecell}

\usepackage{framed}
\definecolor{light-gray}{gray}{0.95}
\renewenvironment{framed}{%
  \MakeFramed {\advance\hsize-\width \FrameRestore}}%
 {\endMakeFramed}

\begin{document}

\title{From Crafting Text to Crafting Thought: Grounding AI Writing Support to Writing Center Pedagogy}
\newcommand{\system}{\textit{Writor}}

\author{Yijun Liu}
\orcid{0009-0008-6601-9237}
\affiliation{%
  \institution{University of Illinois Urbana-Champaign}
  \city{Urbana}
  \state{IL}
  \country{USA}
}
\email{yijun6@illinois.edu}

\author{John Gallagher}
\orcid{0000-0001-5980-4335}
\affiliation{%
  \institution{University of Illinois Urbana-Champaign}
  \city{Urbana}
  \state{IL}
  \country{USA}
}
\email{johng@illinois.edu}

\author{Sarah Sterman}
\orcid{0000-0002-9282-559X}
\affiliation{%
  \institution{University of Illinois Urbana-Champaign}
  \city{Urbana}
  \state{IL}
  \country{USA}
}
\email{ssterman@illinois.edu}

\author{Tal August}
\orcid{0000-0001-6726-4009}
\affiliation{%
  \institution{University of Illinois Urbana-Champaign}
  \city{Urbana}
  \state{IL}
  \country{USA}
}
\email{taugust@illinois.edu}

\begin{abstract}
As AI writing tools evolve from fixing surface errors to creating language with writers, new capabilities raise concerns about negative impacts on student writers, such as replacing their voices and undermining critical thinking skills. To address these challenges, we look at a parallel transition in university writing centers from focusing on fixing errors to preserving student voices. We develop design guidelines informed by writing center literature and interviews with 10 writing tutors. We illustrate these guidelines in a prototype AI tool, Writor. Writor helps writers revise text by setting goals, providing balanced feedback, and engaging in conversations without generating text verbatim. We conducted an expert review with 30 writing instructors, tutors, and AI researchers on Writor to assess the pedagogical soundness, alignment with writing center pedagogy, and integration contexts. We distill our findings into design implications for future AI writing feedback systems, including designing for trust among AI-skeptical educators.
\end{abstract}




\begin{teaserfigure}
\centering
\includegraphics[width=0.84\textwidth]{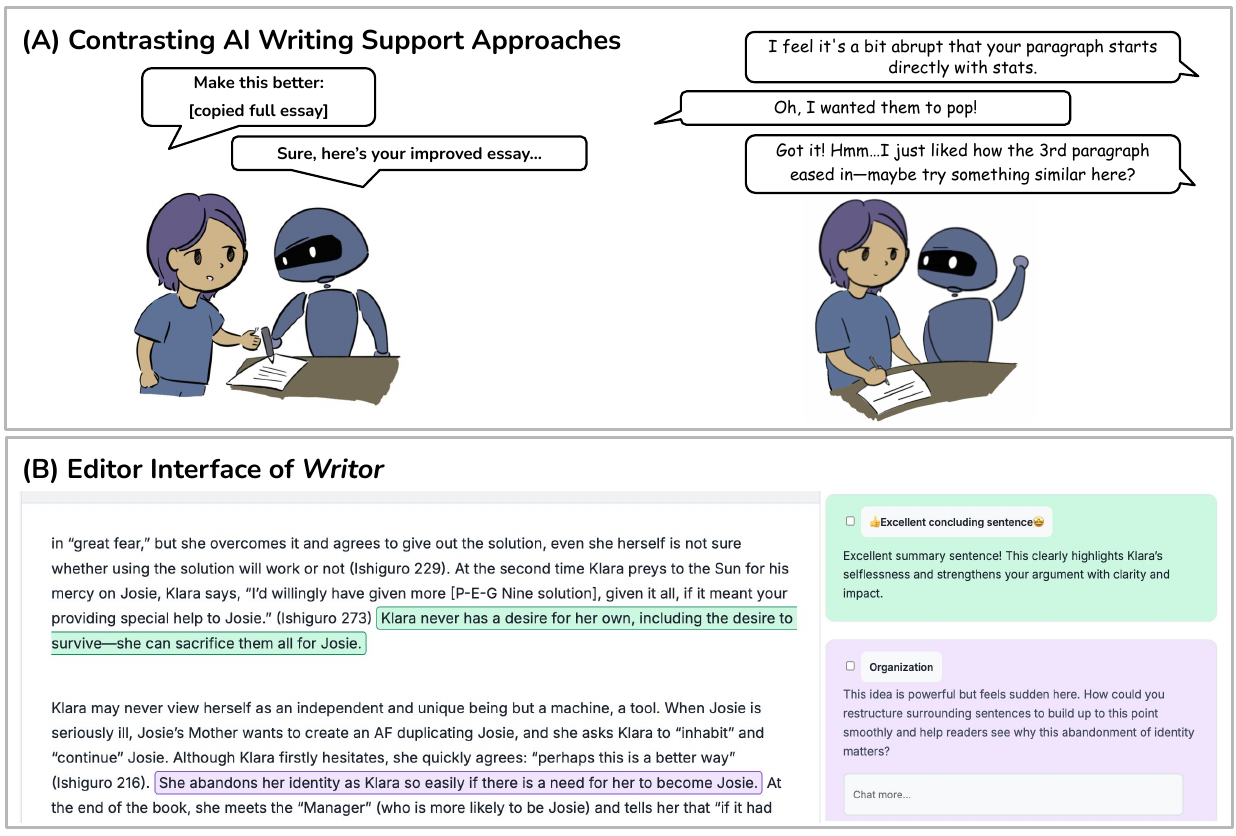}
\caption{Chat-based LLMs typically generate or rewrite text for direct use, Writor offers process-focused, non-directive feedback grounded in writing center pedagogy. }
\Description{The figure includes two parts, contrasting AI writing support and showing the system developed in the paper. The first part contrasts AI writing support approaches: on the left, a chat-based AI rewrites an entire essay for the user; on the right, our system, Writor, engages in a conversational exchange, offering non-directive feedback and suggestions. The second part shows Writor’s editor interface with highlighted sentences and side-panel comments giving targeted, non-directive writing feedback (including both critiques and praises) and prompts for revision.}
\label{fig:writor_overview}
\end{teaserfigure}

\maketitle

% DONE final checking 

\section{Introduction}

The integration of Large Language Models (LLMs) into writing workflows has introduced a tension in Human-Computer Interaction (HCI): how to scaffold writing processes when systems can produce ready-to-use text easily. While commercial and research systems alike have expanded their capabilities from checking grammar to helping writers with their writings' flow, coherence, and audiences-awareness \cite{Ding2024}, their interaction paradigms largely rely on direct text generation and replacement. By functioning as ``proxies'' that perform writing tasks \textit{for} the user, these tools prioritize immediate task completion over the writer's development, thereby drawing on an ideology of workplace efficiency. This design approach risks turning AI writing assistants into ``fix-it shops'' due to their ability to generate quick textual edits, consequently raising questions within the HCI community about how to design systems that support, rather than displace, the human in the writing process.

To navigate this tension, we look to a historical precedent in the evolution of writing support: the university writing center.
Around the 1940s, writing centers, then referred to as writing ``labs'' or ``clinics,'' began emerging at universities across the United States, aimed at addressing student writing problems \cite{carino1995early, carino1996open}. As with many automated writing assistants today, these centers functioned as fix-up shops, ``cleaning up'' students' papers before submission \cite{carino1995early, north1984idea}. Forty years later, the purpose of writing centers began to shift. \textbf{Following the emergence of the process movement in Writing Studies (1970s-1980s), writing centers became more writer-centered rather than curriculum-centered, process-oriented rather than product-oriented, and collaborative rather than didactic in terms of the student-tutor relationship} \cite{carino1995early, north1984idea}. In response to growing concerns of student ownership and plagiarism amid these changes, writing centers developed strategies that realized the vision of collaborative writing spaces while centering the student within the writing process \citep{brooks1991minimalist, 2012oxford, manning2012synthesis}. Writing centers have become a widely adopted, effective resource across institutions worldwide, thereby demonstrating the value of collaborative writing support \cite{archerAfrica, tanWCAsia, boquetIdeaofWC}. They improve student self-efficacy \cite{lundin_impact_2023}, enhance writing quality \cite{salazar_meaningful_2021, zuccarelli_measuring_nodate}, and encourage sustained engagement, with students frequently returning for continued support \cite{bromley_l2_2018}. These outcomes highlight that the strength of writing support may lie less in textual correction and more in framing writing as a social process that values iteration through collaboration. 

In this paper, we operationalize these pedagogical principles for scaffolding writing processes and show potential ways such tools can be incorporated into existing writing support ecosystems.
We thus ask the question, \textbf{how can AI writing tools be designed to facilitate writer-focused, process-oriented, and collaborative writing sessions?} 

We answer this question through a three-stage research process. First, we conducted formative interviews (\S\ref{sec:formative}) with writing tutors ($N=10$) to identify pedagogical practices that we then integrated into concrete design guidelines (\S\ref{sec:designgoals}). Second, we used these design guidelines to develop \system{}, an LLM-based prototype that emphasizes non-directive feedback by restraining copyable text output (\S\ref{sec:system}). The prototype, in sum, does not write for users but instead offers writing-center-like feedback. Finally, to validate this approach, we gathered feedback from 30 domain experts, i.e., writing instructors, writing center tutors, and AI researchers, through a structured survey and follow-up interviews (\S\ref{sec:study2}).

This expert pedagogical review yielded two critical findings regarding the viability of our design approach. First, experts perceived \system{}'s non-directive feedback approach as equivalent or superior to commercial chat-based LLMs, with the strongest endorsement coming from educators skeptical of AI in writing. Second, while experts confirmed that \system{} does not replicate the depth of human instruction, they verified its ability to align with pedagogical principles, specifically in maintaining a positive tone in critiques, offering praises, and preserving writer agency. The review also mapped out integration contexts where tools like \system{} can complement the existing human writing support ecosystem, such as preparing students for tutoring sessions or modeling peer review. We further derive implications from the findings that can be applied broadly in HCI.

In the paper, we make the following contributions:

\begin{enumerate}
    \item We propose using writing center pedagogy as a theoretical foundation for AI writing support tools, focusing on generating non-directive feedback rather than text that can be directly copied.
    \item Through interviews with writing tutors ($N=10$), we translated writing center pedagogy into seven design guidelines for AI writing tools, bridging educational theory and system design.
    \item We designed and developed \system{}, a prototype that operationalizes these guidelines and emphasizes non-directivity that facilitates writing processes.
    \item We report review and feedback from domain experts (writing instructors, writing center tutors, and AI researcher, $N=30$) to validate \system{}'s translation of pedagogical design guidelines, pedagogical soundness, and potential integration contexts for \system{} to supplement the existing writing support ecosystem. 
\end{enumerate}

% DONE final checking 

\section{Related Work}
\label{sec:RW}

\subsection{Writing with AI}

Many users engage directly with chat-based foundational language models like ChatGPT\footnote{\url{https://chatgpt.com/}}, Claude\footnote{\url{https://claude.ai/}}, and Gemini\footnote{\url{https://gemini.google.com/}} for various writing tasks \cite{marzuki_impact_2023, Ding2024, dwivedi_opinion_2023, minaSpace}. These general-purpose conversational systems have become popular writing companions, with users employing them for brainstorming ideas \cite{mahapatra_impact_2024, pokkakillath_chatgpt_2023, koos_navigating_2023, dergaa_human_2023, wansecondmind, di_fede_idea_2022, sparkkaty}, generating initial drafts and summaries \cite{dwivedi_opinion_2023, pokkakillath_chatgpt_2023, dergaa_human_2023, storey_ai_2023, jelson_empirical_2025, tica2024overcoming}, refining writings \cite{dwivedi_opinion_2023, koos_navigating_2023, xu_patterns_2025, storey_ai_2023, jelson_empirical_2025}, and providing feedback on existing text \cite{viantika_improving_2024, tran_enhancing_2025, guo_resist_2024, nazari_application_2021, leoste_perceptions_2021, steiss_comparing_2024}. Users often treat these models as writing partners, bouncing ideas back and forth \cite{marzuki_impact_2023, pokkakillath_chatgpt_2023, jelson_empirical_2025}, exploring different approaches to structure and tone \cite{mahapatra_impact_2024, koos_navigating_2023}, and using the conversational format to engage and think through complex arguments together \cite{marzuki_impact_2023, song_enhancing_2023, jelson_empirical_2025}. In addition to chat, many tools now integrate dedicated ``canvas'' writing modes, enabling inline editing and rewriting alongside the chat interface \cite{kannan2025canvas}. 

Similar to the shift that occurred with writing centers, there has also been growing interest in specialized writing tools to support writers as collaborators or to provide feedback for revisions. A common theme in current AI writing assistants is to treat the model as a co-author \citep{wordcraft_yuan, mirowski_co-writing_2023}. This idea extends across domains, such as personal journaling \citep{kim_diarymate_2024}, academic review writing \cite{su-etal-2023-reviewriter, choe_supporting_2024}, legal writing \cite{legalwrite}, creative stories \cite{creative_writing_elizabeth, creative_rnn, ghajargar_redhead_2022}, and creating characters \cite{qin_charactermeet_2024}. Another way to support writers as co-authors is emphasizing process guidance and exploration over content generation. This approach spans contexts ranging from structured, step-by-step support to interactive environments that prompt reflection and iteration \cite{choe_supporting_2024, khan2024writingcoach, talaei_storysage_2025, arnold_interaction-required_2025}. In addition to guiding users through multiple stages of drafting and revision, these systems encourage exploration of alternative phrasings \cite{reza_abscribe_2024}, metaphors \cite{Gero19metaphoria}, and collaboration paradigms  \cite{siddiqui_scriptshift_2025}. 

A second theme is to provide feedback for revising drafts. Many tools provide feedback from surface-level suggestions to persona-based feedback \citep{writer_persona}. Surface-level tools like Grammarly\footnote{\url{www.grammarly.com}} focus on grammar and spelling corrections, while more sophisticated systems address broader structural adjustments \cite{legalwrite, llm_classroom_meyer, han2024llmasatutoreflwritingeducation, bella_rhetor}, language enhancement \cite{empathy_learning, llm_classroom_meyer, han2024llmasatutoreflwritingeducation}, and adherence to writing requirements \cite{dai2023can, han2024llmasatutoreflwritingeducation}. These feedback systems employ diverse techniques to support writers. Some use Socratic questioning to prompt deeper reflection and critical thinking \cite{kim2023repurposingtextgeneratingaithoughtprovoking, arnold2021generative, choe_supporting_2024}, while others provide continuous summaries to help writers reflect on their writing processes \cite{10.1145/3526113.3545672}.
Reflection can also be supported through interface-level nudges, such as analogies, personas \cite{yeo_help_2024}, reflection timers \cite{yeo_not_2024}, and delays \cite{bae_ripening_2014}.
Advanced systems attempt to align feedback with a writer's values and intentions, ensuring contextual relevance \cite{kreminski-martens-2022-unmet}. Recent systems such as Friction visualize feedback across a draft to help writers prioritize issues and plan actionable revisions \cite{zhang_friction_2025}.

While existing AI writing systems demonstrate diverse approaches to supporting writers, most systems operate through text generation for users (i.e., either by directly editing text or making suggestions). This directive guidance (e.g., making a wording suggestion change) can diminish writer agency \cite{reza2025cowritingaihumanterms} and has led to many educational concerns (\S\ref{sec:risingConcerns}) even as AI writing tools are fast being adopted. \textit{Our work aims to reconcile the growing demand for AI writing tools with educator concerns by drawing from writing center pedagogy.} \looseness=-1

\subsection{Rising Concerns in Writing with AI}
\label{sec:risingConcerns}

A central concern of writing with AI is that it challenges traditional notions of authorship and originality \cite{dwivedi_opinion_2023, miao_ethical_2023, rainie_watson_2025, hwang_it_2024}, raising widespread fears around plagiarism and academic integrity \cite{rainie_watson_2025, santiago_utilization_2023, malik_exploring_2023, semrl_ai_2023, cotton_chatting_2023, almaiah_measuring_2022}. As LLMs can generate coherent, evidence-rich content with minimal user input, they also complicate questions of agency and ownership---how writers perceive their connection to and control over their work when they write with LLMs \cite{katy_creative}. Research has shown that users' sense of agency and ownership---both intellectual and emotional---can be diminished depending on how AI systems are designed \cite{reza2025cowritingaihumanterms, Stark2023, 10.1145/3698061.3726907}. For instance, longer generated text and direct idea generations often reduce perceived control and ownership \cite{robertson_i_2021, shaping_scaffolding_dhillon, draxler_ai_2024, katy_creative, biermann_tool_2022}, 
while interaction design such as provenance tracking can restore agency \cite{hoque_hallmark_2024}.

From an educational standpoint, AI can hinder writing skill development. Students may bypass critical thinking and revision processes \cite{marzuki_impact_2023, kosmyna_your_2025, dong_revolutionizing_2023}. There are also worries that habitual reliance on AI for brainstorming or drafting will weaken creativity and problem-solving capacities \cite{dwivedi_opinion_2023, mahapatra_impact_2024, zhai_effects_2024, koos_navigating_2023}. AI responses can also homogenize student viewpoints, as outputs tend to reflect dominant patterns rather than diverse or critical perspectives \cite{marzuki_impact_2023}. Moreover, users can develop over-reliance on the tools, amplifying these risks \cite{marzuki_impact_2023, dwivedi_opinion_2023, mahapatra_impact_2024, song_enhancing_2023, zhai_effects_2024}.

AI suggestions can also shape users’ stances on controversial topics \cite{jakesch_co-writing_2023}, push writing toward dominant cultural norms \cite{agarwal_ai_2025}, and guide content selection in self-presentation \cite{poddar_ai_2023}, yet this influence often goes unnoticed by users. Another major concern is accuracy: language models are prone to hallucinating facts or generating misleading information \cite{watts_comparing_2023, santiago_utilization_2023, malik_exploring_2023, hellstrom_ai_2024, xu_patterns_2025, koltovskaia_graduate_2024}. Additionally, AI-generated content often exhibits formulaic linguistic patterns \cite{mahapatra_impact_2024}, frequently lacking the originality and conceptual novelty characteristic of human-authored work \cite{dwivedi_opinion_2023, koos_navigating_2023}.

While challenges such as factual inaccuracy and algorithmic bias in LLMs raise broad ethical concerns that require systemic and technical solutions, other issues---such as plagiarism, over-reliance, diminished creativity, and reduced writer agency---are more directly shaped by how AI writing tools are designed. \textit{In this paper, we rethink the form of AI writing support: moving away from usable text generation toward pedagogically grounded strategies that encourage writer agency.} \looseness=-1

\subsection{Writing and Writing Center Pedagogy}
\label{sec:WC_lit}

Current writing center pedagogy has been largely influenced by \citet{north1984idea}'s seminal essay ``The Idea of a Writing Center'' in 1984. Since then, these ideas have evolved into more specific strategies. This section synthesizes writing center pedagogy into three distinct, interconnected themes of writing support: \textbf{1) writer-centered, 2) process-oriented, and 3) collaborative.} 

The \textit{writer-centered approach} tailors support to ``the writers it serves'' rather than to fixed curricula \cite{north1984idea}. To foster a writer-centered environment, writing center literature has developed a set of individualized scaffolding techniques. One key type of scaffolding is motivational scaffolding, which aims to cultivate students' interests in writing tasks and encourage their persistent engagement with writing processes \cite{2012oxford, mackiewicz2013motivational, cromley2005reading, wood1976tutoring}, through practices such as offering specific praises and demonstrating sympathy and empathy \cite{mackiewicz2013motivational}. 
\looseness=-1

A \textit{process-oriented approach} to writing instruction emphasizes developing writers' skills over writers' texts \cite{north1984idea}. One influential framework within this approach is minimalist tutoring. Initially focused on promoting student ownership of their work \cite{braun2006using}, minimalist tutoring later evolved into a widely accepted strategy for fostering student learning by minimizing direct intervention \cite{2012oxford}. Instead of providing students with explicit corrections, minimalist tutoring encourages them to engage actively in the writing process \cite{brooks1991minimalist}. Ultimately, the goal of minimalist tutoring is to cultivate independent writers who can critically assess and refine their work \cite{2012oxford}. Writing center literature reinforces this process-oriented approach through several scaffolding techniques to give feedback that encourages students to engage with their writing \cite{2012oxford}. Key scaffolding strategies include reacting as a reader, where tutors provide feedback from the perspective of an imagined reader and metacommentary, where tutors explain the reasoning behind feedback to help students internalize the revision process \cite{2012oxford}.

A \textit{collaborative approach} in writing centers emphasizes partnership between tutors and writers rather than a hierarchical instructional model \cite{north1984idea}. Instead of tutors simply directing students, both parties engage in dialogue. Writing center literature shows that collaborative approaches encourage critical thinking \cite{bruffee1984peer} and deeper engagement with writing \cite{manning2012synthesis, thompson2009scaffolding}. Through discussion, tutors provoke thought in a social context, encouraging active learning \cite{bruffee1984peer}. In writing center literature, writing itself is often viewed as a re-externalized conversation, meaning that the writing process mirrors the way ideas are developed and refined through dialogue \cite{bruffee1984peer, mcandrew2001tutoring}. By engaging in dialogic interactions about their writing, students can improve their ability to articulate ideas clearly, refine their arguments, and thereby develop stronger writing skills \cite{2012oxford}. 

Although writing center pedagogy offers well-established strategies to support student agency, collaboration, and process-based learning, its insights are rarely integrated into the design of AI writing tools. \textit{Our work bridges this gap by systematically translating writing center pedagogy into actionable design guidelines (\S\ref{sec:designgoals}) and providing an example of their implementation with a prototype LLM-based writing tool, \system{} (\S\ref{sec:system}).}

% DONE final checking 

\section{Formative Interview}
\label{sec:formative}

To ground the writing center principles introduced in \S\ref{sec:RW} into design guidelines for intelligent writing support, we examined how writing tutors implement them in tutoring sessions through semi-structured interviews. While writing center literature outlines pedagogical principles, their implementation is deeply contextual and situated. Talking directly with tutors allowed us to see how they translate abstract principles into concrete strategies and to explore their perspectives on current AI's role in writing centers.

The study was guided by two research questions: \textbf{(1) What strategies do writing tutors use to support students, and how are these strategies implemented during tutoring sessions? (2) What are tutors' perspectives on how AI could support or transform writing center practices?}

\subsection{Procedure}
\label{sec:interview_procedure}

We conducted semi-structured interviews with writing center tutors. Interested tutors completed an online screening survey to confirm their tutoring background (have worked in a writing center for at least three months). All interviews were conducted in English via Zoom, lasted approximately 60 minutes, and each participant received a \$20 Amazon gift card as compensation. Interviews were audio-recorded, transcribed, and anonymized. The interview schedule is included in the Appendix \ref{sec:interview_guide}. This study was approved by the relevant Institutional Review Board.

\subsection{Participants}

Participants were recruited through multiple channels, including writing center email lists, campus fliers, and participant referrals. Fliers were posted on a college campus, and recruitment emails were sent to writing center administrators at U.S. universities, who were asked to share the study information with their tutoring staff. We also employed snowball sampling, where tutors who agreed to participate could refer other tutors they knew who may be interested by sharing the interest form. 

We recruited ten writing center tutors from three U.S. universities, including two private institutions and one public institution. As writing centers operate on a peer-tutoring model, the sample included six graduate students and four undergraduate students. Participants’ tutoring experience ranged from three months to five years, with an average of 1.95 years (SD = 1.54 years). 

\subsection{Analysis}

To identify themes and strategies tutors used to support student writing processes, we conducted a reflexive thematic analysis on the transcribed interviews following ~\citet{braun2006using}. Although existing writing center literature provided concrete theoretical grounding, we chose this approach over purely deductive coding to remain open to tutors' situated practices that may not align with established categories and to surface emergent themes around how principles are being adapted in the age of AI. 

After conducting the interviews, the first author familiarized herself with the interview data and made initial notes on tutoring strategies and themes. This author created an initial set of codes for individual strategies (e.g., ``providing reader-perspective feedback'') and iterated on these codes through discussions with the last author. Iteration happened weekly during in-person discussions over the course of a month and included the last author and first author recoding the same interview transcript and meeting to resolve differences in codes. Following iteration, the authors reviewed the strategies and transcripts collectively to assess supporting evidence for each strategy. After refining the strategies, the first author revisited the data and checked for consistency between strategies and observations from the study. 

Throughout the analysis, we iteratively engaged with writing center literature and guidelines to understand how emergent strategies connected to established pedagogical principles. In the following section, we describe the strategies surfaced by our interviews, organized around the writing center literature's characterization of support as \textit{writer-centered}, \textit{process-oriented}, and \textit{collaborative}. 

\paragraph{Positionality} Both coders (first and last authors) previously trained and served as writing center tutors and now work as AI/HCI researchers. Their tutoring experience provided firsthand insight into writing center pedagogy, which allowed them to quickly connect participant comments to established writing center literature. Many of the strategies identified therefore resonated as confirmatory of existing practices. At the same time, their roles as AI/HCI researchers meant they approached the data with an interest in how AI might authentically embody these pedagogical values.

% DONE final checking 

\section{Formative Interview Findings}

\subsection{Writing Support is \textit{Writer-Centered}}

\subsubsection{Empathy and Building Confidence.}
\textit{``How do you feel?''} Six out of ten tutors mentioned this specific phrase during their interviews as they described what they would typically say to students. Empathy and confidence building emerged as an important writer-centered approach, often mentioned as a means of \textit{motivational scaffolding} in writing center literature \cite{2012oxford, mackiewicz2013motivational}. For example, P5 mentioned listening to and reassuring students when they felt frustrated with reviewer comments or their relationship with advisors, while P7 deliberately tried to build emotional rapport to help students feel more comfortable during sessions. 

This foundation of empathy naturally fed into confidence-building, where tutors used encouraging language and praise to help students recognize their own progress. Five tutors used encouraging language and verbal compliments to affirm students’ writing abilities. For example, P2 emphasized the importance of helping students recognize their own progress, creating an environment where the student can believe that, \textit{``Yes, [student] can be a writer''}; P5 described boosting students' confidence by reassuring them that their writing was already strong, particularly for those experiencing imposter syndrome or writing in non-native languages.

\subsubsection{Preserving Students' Voices.}
Another crucial aspect of the writer-centered approach was preserving students' voices in writing. Six tutors emphasized the importance of maintaining students' original meaning and personal characteristics in their writing. As P2 noted, they prefer to \textit{``keep [students' writings] as a kind of personal characteristic.''} P8 highlighted their training to emphasize \textit{``it's the students' ideas that we're working with''} rather than imposing their own thoughts. P3 also employed a strict rule in giving students no more than four continuous words to ensure this, because \textit{``sometimes I say a sentence, and they [students] go: `Oh, that's what I like.''} This focus on preserving student voices and maintaining ownership aligns with many centers' minimalist, non-directive tutoring \cite{manning2012synthesis}. 

\subsubsection{Centering the Writer with AI}
Five tutors brought up issues related to plagiarism raised by supervisors or students when discussing AI in tutoring contexts. P1 described a common scenario: \textit{``You [the student] use the words correctly, but you don’t know what they mean.''} Tutors argued that any AI writing support should prioritize student voices by adopting non-directive, minimalist feedback. As they explained, such an approach should \textit{``force people to make sure what they[AI] generates'' (P5)} and \textit{``tell them[AI] not to touch any content'' (P8)}.

\begin{figure*}[t]
    \centering
    \includegraphics[width=0.76\textwidth]{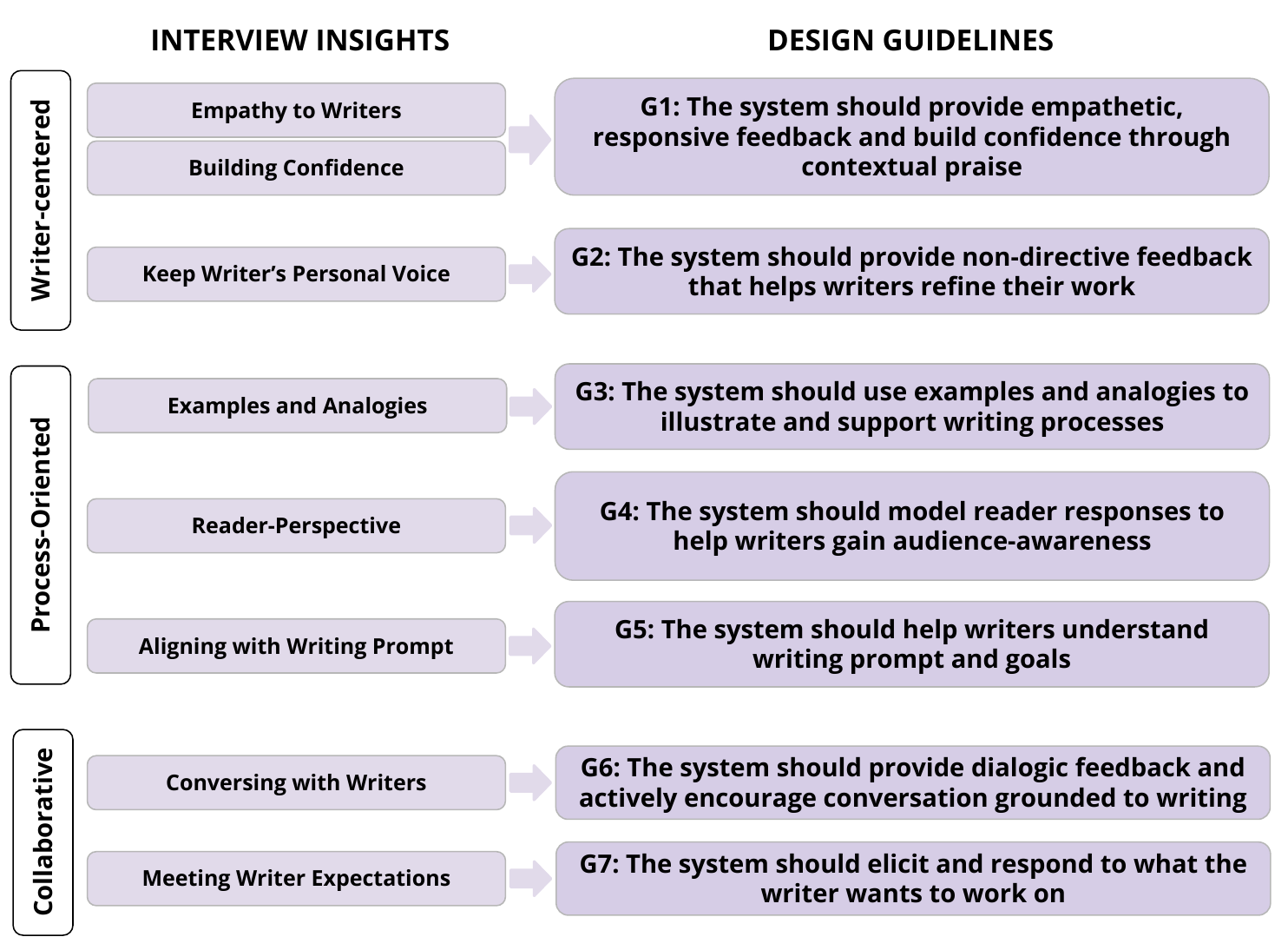}
    \caption{ Interview insights and design guidelines categorized by writer-centered, process-oriented, and collaborative. Based on insights from formative interviews with writing tutors, we identified seven core design guidelines.}
    \Description{A diagram with two columns: Interview Insights on the left and Design Guidelines on the right, connected by arrows. The rows are grouped into three categories. Writer-centered: Empathy to Writers and Building Confidence link to G1, "The system should provide empathetic, responsive feedback and builds confidence through contextual praise"; Keep Writer's Personal Voice links to G2, "The system should provide non-directive feedback that helps writers refine their work." Process-Oriented: Examples and Analogies link to G3, "The system should use examples and analogies to illustrate and support writing processes"; Reader-Perspective links to G4, "The system should model reader responses to help writers gain audience-awareness"; Aligning with Writing Prompt links to G5, "The system should help writers understand writing prompt and goals." Collaborative: Conversing with Writers links to G6, "The system should provide dialogic feedback and actively encourage conversation grounded in writing"; Meeting Writer Expectations links to G7, "The system should elicit and respond to what the writer wants to work on."}
    \label{fig:guidline}
\end{figure*}

\subsection{Writing Support is \textbf{\textit{Process-Oriented}}}

Process-oriented writing support strategies emerged as another significant theme in our study. We identified three key strategies, with each being independently mentioned by eight tutors during our interviews. 

\subsubsection{Using Examples and Analogies.}
Tutors emphasized the use of examples and analogies to facilitate student learning and comprehension. They provided a wide range of examples, from providing sentence structure options to sharing personal experiences for understanding writing contexts. They also used analogies to clarify complex concepts. For instance, P3 described using a simple topic---such as apples---to illustrate how to structure an introduction: \textit{``If I was writing a paper on apples, I would start with a broader history of apples and how they fit into my thesis, and then gradually lead into the thesis itself.''}

\subsubsection{Providing Reader-Perspective.}
Eight tutors delivered feedback from a reader's perspective rather than a purely instructional standpoint. Instead of providing directive feedback as tutors, they shared their reactions and understanding of the text as readers, helping students recognize how their writing affects their audience. For example, P9 provided their perspective as a reader and asked clarifying questions accordingly by asking questions like, \textit{``I also noticed [something] as I was reading...maybe you could expand here?''}

\subsubsection{Understanding Prompts.} 
Tutors ensured students thoroughly understood assignment prompts to maintain alignment with instructor expectations. Specifically, this involved having students explain prompts in their own words and collaboratively reading instructions to establish shared understanding. For instance, P6 emphasized checking for misalignment, noting that if students explained the instructions differently from how they initially interpreted them, they would \textit{``literally point to parts of the instructions and say, ‘When your instructor says to add more ethos to your paper, what does ethos mean?’''}

\subsubsection{Orienting Process in AI} 
Organization and planning was the most highly regarded capability of AI tools among tutors. Tutors shared their experiences using AI for organization, flow, generating ideas, and outlining. P3 imagined a scenario where a student might ask how their essay differs from a standard academic essay, and that an AI model could be helpful in simply pointing out basic elements, such as telling a freshman, \textit{``you need an introduction.''} Three tutors reported using AI for these purposes in their own writing, while one tutor had experience using AI for writing tutoring. In total, six tutors perceived that AI could be effectively employed in writing tutoring for process-oriented tasks, making it the most highly perceived capability of AI.

\subsection{Writing Support is \textit{Collaborative}}

\subsubsection{Tutoring is Conversational.}
\textit{``Sometimes they show the writing... I say, close your laptop for a second, and I close my laptop, too, and I say, like, talk to me,'' } P6 said. Six tutors highlighted how conversation forms the foundation of collaborative writing support. Tutors actively created opportunities for dialogue rather than delivering one-way instruction. Some tutors, like P6 and P8, intentionally asked students to close their laptops to facilitate conversation, shifting focus from the written text to the verbal expression of ideas. Tutors also engaged in back-and-forth discussions instead of direct instruction about writing strategies and clarity, as noted by P4 and P9. Additionally, P7 emphasized that conversation is central to their tutoring philosophy, using dialogue as a primary tool for helping students develop their ideas. 

\subsubsection{Understanding Expectations.}
Rather than prescribing solutions, five tutors described a collaborative process of understanding writers' expectations. Tutors usually dedicated initial session time to mutual exploration of achievable goals for the session that are \textit{``the most helpful to [the writer]''}, as P7 noted. This approach positioned writers as active participants rather than passive recipients of instruction, creating a shared understanding that guided their collaborative work rather than imposing a tutor-directed agenda. \textit{``[The way] we're trained isn't necessarily like we're gonna go through and tell you everything,''} P9 emphasized. 

\subsubsection{Collaboration in AI Systems} While tutors did not specifically mention AI's capabilities in fostering a similar collaborative writing space, five tutors expressed skepticism about AI’s ability to truly understand student writing. They noted that tools like ChatGPT often provide overly general feedback, struggle with long or complex texts, and lack awareness of context or instructional expectations. Unlike human tutors who facilitated back-and-forth conversations, AI systems seemed only capable of engaging in vague discussion. As P6 noted, \textit{``ChatGPT sounds like the classmate who didn't do the reading, but still has to participate in class.''}

\subsection{Design Guidelines}
\label{sec:designgoals}

We translate tutoring practices from our interviews into actionable design guidelines for intelligent writing support systems, as shown in Figure \ref{fig:guidline}. We unified guidelines around how we perceived these strategies interacting based on interviews with tutors and writing center literature. ``Empathy to Writers'' and ``Building Confidence'' were merged into G1, as both contribute to motivational scaffolding.

\section{System Design}
\label{sec:system}

We illustrate how our design guidelines can inform the design of intelligent writing tools by developing \system{}, a prototype writing system that aims to provide process-oriented, writer-centered, and collaborative writing support. 

\subsection{System Architecture \& Workflow}
\label{sec:architecture}

\system{} utilizes a client-server architecture with a Flask-based backend and a JavaScript frontend. The backend integrates prompting to OpenAI's GPT-4.1-mini for writing analysis and Firebase Firestore to store session data and interaction history.

The system operates in two key stages: (1) a preparation and goal-setting stage, where writers input context, upload texts, and define their writing objectives (\S\ref{sec:planning}); and (2) an editing stage, where they receive and engage with AI-generated feedback (\S\ref{sec:editing}). Prompts used for \system{} are included in Appendix \ref{sec:appendix}.

\subsection{Preparation \& Goal Setting}
\label{sec:planning}

\begin{figure}[t]
    \centering
    \includegraphics[width=\columnwidth]{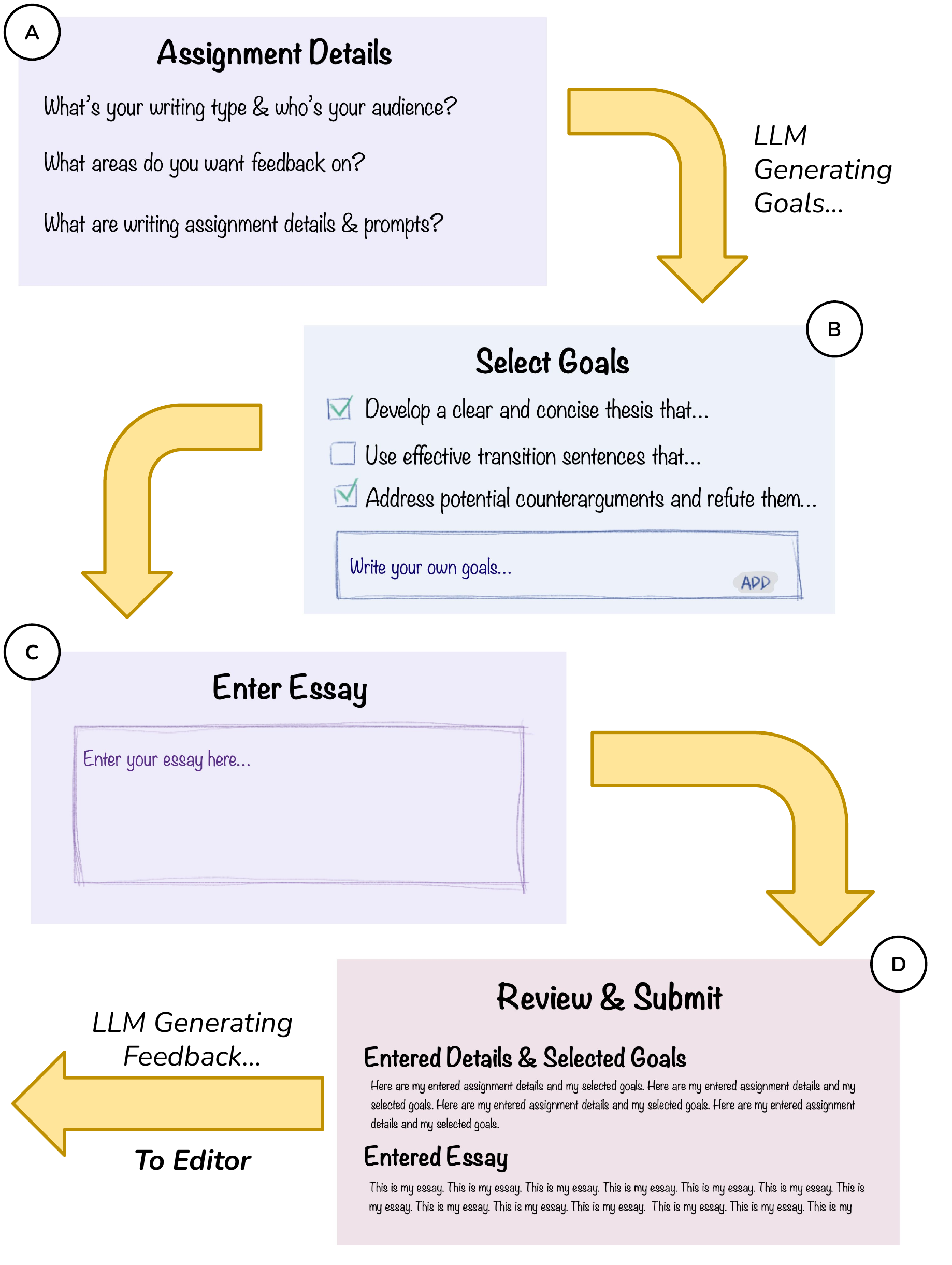}
    \caption{Illustration of Preparation \& Goal Setting Stage of \system{}: (A) entering writing task’s purpose and requirements, (B) selecting goals to focus on during revision, (C) enter the essay, and (D) review the input and selected goals. See Fig. ~\ref{fig:interface-goal-real} in the appendix for actual system interfaces.}
    \Description{An illustration of the interfaces for the Preparation and Goal Setting stage of Writor. Four labeled panels show: (A) assignment details page that includes text boxes to enter the writing's audience, tasks, and areas needing feedback, (B) select goal page a list of selectable goals to focus on during revision and an option to enter goals, (C) a large input area to paste or type the essay, and (D) a summary screen that reviews both the entered text and the chosen goals, and after this page, users will be directed to the editor page with generated feedback. LLMs generate the goals from the entered assignment details.}
    \label{fig:interface-goal}
\end{figure}

Two tutors in our study noted that students often struggle to articulate goals independently or propose goals misaligned with writing center practices (e.g., seeking only grammar correction). Because of this, \system{} generates goals as starting points for writers to think through their own goals. 

Writers start by reflecting on three planning questions (Fig. \ref{fig:interface-goal} A): their task details,  potential readers, and areas they want to improve. Based on the writer's input, \system{} generates a list of five suggested goals (Fig. \ref{fig:interface-goal} B, \textbf{G5}), with one of these goals focusing on potential audiences. An example of a generated goal is ``use effective transition sentences and phrases to ensure smooth and logical flow between paragraphs and ideas.'' Writers can then select which of the generated goals to use and write in their own additional goals (\textbf{G7}). After setting the goal, writers will then input their essay and review their inputs before going to the editor interface. Figure \ref{fig:interface-goal} shows a high-level illustration of the goal-setting stage, and screenshots for these four stages can be found in Appendix Fig. \ref{fig:interface-goal-real}.

\subsection{Editing Stage}
\label{sec:editing}

\begin{figure*}[t]
    \centering
    \includegraphics[width=\textwidth]{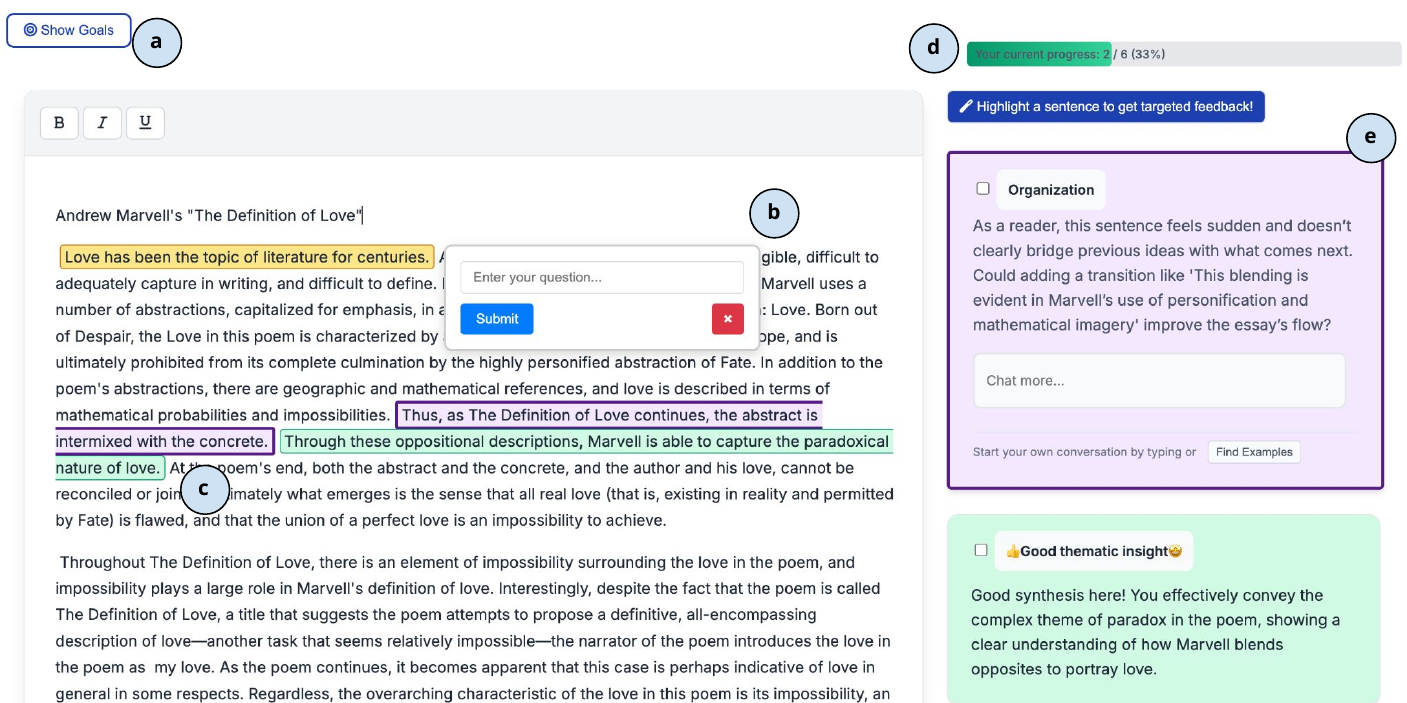}
    \caption{Edit Interface of \system{}: (a) Expandable button for viewing selected writing goals; (b) Highlight \& Get Feedback for user-initiated analysis; (c) Interactive highlighting that connects feedback to specific text; (d) Progress bar for tracking addressed feedback items; (e) Critique and praise cards.}
    \Description{Screenshot of the editing interface. Left pane: a rich-text editor with a multi-paragraph essay and sentences highlighted in yellow, purple, and green. A pop-up next to the yellow highlight says ‘Enter your question…’, showing user-initiated targeted feedback. A top-left button labeled ‘Show Goals’ displays goals from the previous stage. Right pane: a progress bar, a blue button ‘Highlight a sentence to get targeted feedback!’ linked to the yellow highlight, a purple feedback card labeled ‘Organization’ with guidance text and a chat box, and a green praise card titled ‘Good thematic insight’. Feedback card colors correspond to the highlighted text.}
    \label{fig:interface}
\end{figure*}

Following goal-setting, \system{} automatically overlays feedback (Fig. \ref{fig:interface}) aligned to the writer's goals on the writer's working draft. There are two types of feedback: critique and praise. Below we describe the individual interface elements that support the editing stage.

\subsubsection{Text Editor}
The text editor on the left panel serves as the core space for writing and revising. It provides basic text formatting tools including bold, italic, and underline. The two key features in the editors are:

\begin{itemize}
    \item \textbf{Interactive Highlighting (Fig. \ref{fig:interface}c)}: Each suggestion or comment generated by \system{} is linked directly to highlighted spans of the writer's text, color-coded to critique or praise. When a highlighted span is clicked, the associated feedback card will scroll into view in the sidebar. 
    
    \item  \textbf{Highlight \& Get Feedback Button (Fig. \ref{fig:interface}b)}: \system{} allows writer-initiated feedback in addition to initial generated feedback (\textbf{G7}). A writer can highlight any sentence within the text editor and request feedback from \system{}. 
\end{itemize}

\subsubsection{Feedback Sidebar}
\label{sec:feedback}

\system{}'s sidebar (Fig. \ref{fig:interface}e) holds cards representing all generated feedback, including praise \textbf{(G1)} and areas for improvement. For each area of improvement, \system{} is designed to generate feedback using one or more of the following strategies \textbf{(G2)}: asking questions \textbf{(G6)}, providing examples and analogies \textbf{(G3)}, or offering a reader perspective \textbf{(G4)}. We limit initial feedback to five items and three praises in order to not overwhelm writers with long lists of generated feedback.

Writers can ask follow-up questions for each critique within each card \textbf{(G6)}. If a writer wants examples of how to implement a suggestion, a \textit{Find Example} button below the chat bar enables an additional strategy: finding examples within the writer's own text that might be a first step to addressing the current suggestions \textbf{(G3 \& G6)}. If no examples can be found, \system{} provides analogies or examples on a different topic (e.g., P3's use of a basic topic like apples to illustrate a suggested revision).

\begin{table*}[ht]
\centering
\small
\begin{tabular}{p{4.4cm}p{6.2cm}p{4cm}}
\toprule
\textbf{Feature} & \textbf{Description} & \textbf{Design Guideline} \\
\midrule
\addlinespace[0.5em]
\textit{\textbf{Writer-Centered Features (G1--G2)}} & & \\
\makecell[l]{Praise Cards \\ (Sidebar, Fig.~\ref{fig:interface}e)} 
& Highlights strengths in the draft with specific and contextualized praise. 
& G1: Empathy \& Confidence \\
\makecell[l]{Critique Cards \\ (Sidebar, Fig.~\ref{fig:interface}e)}
& Provides non-directive, empathic critiques aligned with goals set in the goal-setting stage (\S\ref{sec:planning}). 
& G2: Non-directive Feedback \\
\makecell[l]{Prompting \\ (Backend, Appendix~\ref{sec:appendix})} 
& A prompting pipeline explicitly forbids copy-pasteable text and enforce empathic, supportive tone in critiques. 
& G2: Non-directive Feedback \\
\makecell[l]{Interactive Highlighting \\ (Editor + Sidebar, Fig.~\ref{fig:interface}c, e)}
& Color-coded linking between text and cards encourages reflection and prevents direct insertion. 
& G2: Non-directive Feedback \\
\addlinespace[1.5em]

\textit{\textbf{Process-Oriented Features (G3–G5)}} & & \\
Analogies \& Examples
& Feedback uses generated analogies/external examples to clarify feedback. 
& G3: Examples \& Analogies \\
Reader Perspective 
& Provides critiques through phrasing as readers' reactions (e.g., ``As a reader, I feel...'') to increase reader-awareness. 
& G4: Reader-Perspective\\
\makecell[l]{Goal Generation \\ (Goal Setting stage, Fig.~\ref{fig:interface-goal}A\&B)}
& Generates five suggested goals from writer-provided task, audience, and concerns. 
& G5: Align with Goals \\
\makecell[l]{Topic Identification \\ (Backend, Appendix~\ref{sec:appendix})}
& Identifies relevant high-level writing topics (e.g., thesis) based on selected goals as part of the critique prompting pipeline. 
& G5: Align with Goals \\
\addlinespace[1.5em]

\textit{\textbf{Collaborative Features (G6–G7)}} & & \\
\makecell[l]{``Find Example'' Button \\ (At the bottom in critique cards)}
& Prioritizes examples from the writer’s own draft to demonstrate how to address the critiques; if none exist, it generates an illustrative example to guide revision. It also offers an easy starting point for conversations.
& G3: Examples \& G6: Dialogic \\
\makecell[l]{Open-Ended Question \\ (Backend)}
& Ensures all feedback ends with a question to invite dialogue. 
& G6: Dialogic Feedback \\
\makecell[l]{Conversational Chat \\ (In each critique card)}
& Enables multi-turn discussion on specific feedback cards while preserving non-directiveness. 
& G6: Dialogic Feedback \\
\makecell[l]{Custom Goal Input \\ (Goal Setting stage, Fig.~\ref{fig:interface-goal}B)}
& Writers can add their own revision goals beyond AI-suggested ones. 
& G7: Meet Writer Expectations \\
\makecell[l]{Revisit Selected Goals \\ (Fig.~\ref{fig:interface}a)} 
& Displays chosen goals for ongoing reference throughout writing. 
& G7: Meet Writer Expectations \\
\makecell[l]{``Get Targeted Feedback'' \\ (Fig.~\ref{fig:interface}b)}
& Allows sentence-specific, writer-initiated feedback requests.
& G7: Meet Writer Expectations \\
\bottomrule
\end{tabular}
\caption{Overview of \system{} features mapped to pedagogical design guidelines.}
\label{tab:writor-features}
\Description{Overview of system features grouped into three categories: writer-centered features, process-oriented features, and collaborative features. Each row lists a feature and its location, a brief description of what it supports, and the pedagogical guideline it corresponds to.}
\end{table*}

\subsection{Summary of Features}
Table~\ref{tab:writor-features} summarizes the key features of \system{}, grouped according to our three pedagogical themes: writer-centered (G1--G2), process-oriented (G3--G5), and collaborative (G6--G7) support. Features are listed with their location, description, and the specific design guideline they address. This overview illustrates how each design guideline operationalizes into \system{}'s features.

\subsection{Internal Audit}
\label{sec:audit}

To ensure that generated feedback aligned with our intended features and writing center standards, we conducted an internal audit of the generated goals (\S\ref{sec:planning}) and feedback (\S\ref{sec:editing}). Two authors independently reviewed system outputs using a structured coding template with defined criteria (listed in Table \ref{tab:audit_summary}). Initial agreement was substantial ($\kappa=0.79$), and disagreements were resolved through discussion. Full definitions and examples are in Appendix~\ref{sec:audit_questions}.

In \system{}, goals and feedback are generated via a mixture of single prompts and multi-prompt pipelines. Goals are generated via a single prompt, and evaluated for relevance and specificity toward writer's writing prompt details and editing expectations. Feedback includes both critique and praise. Critiques use a 4-stage pipeline: (1) identify feedback topics from goals, (2) locate relevant sentences, (3) select feedback type (e.g., giving examples), and (4) generate feedback. Praise uses single prompt generation. We generated and evaluated 24 goals, 24 topic identifications, 60 critiques, and 36 praises across four texts.
We report the results of the audit in Table~\ref{tab:audit_summary}. The results indicated that \system{} successfully met its intended standards: goals were always relevant, specific, and tailored to the prompts (100\% across criteria); critique feedback was accurate, non-directive, and appropriately mapped to higher-order concerns (mostly greater or equal to 96\%, with feedback-type alignment a bit weaker at 88.33\%); and praise was consistently appropriate, though slightly less specific (91.67\%).

\begin{table*}[ht]
\centering
\small
\begin{tabular}{@{}p{2.8cm}p{6.0cm}p{3cm}@{}}
\toprule
\textbf{Audit Component} & \textbf{Evaluation Criteria} & \textbf{Success Rate} \\
\midrule
\textbf{Goal Generation} & Is this goal relevant to the prompt or edits? & 100.00\% \\
                    & Is this goal specific for the prompt or edits? & 100.00\%\\
                         & Is Goal 5 tailored to the instructor's expectations? & 100.00\% \\
\midrule
\textbf{Critique Feedback} & Does this topic (e.g. thesis, argument) fit into the pre-set categories of high-order writing concerns?& 100.00\% \\
                      & Does the topic align with one or multiple goals? & 96.00\%\\
                           & Is the feedback appropriate for the sentence? & 100.00\%\\
                           & Is there any inaccurate information? & 98.33\% \\
                           & Does the feedback include any usable text? & 96.67\% \\
                           & Is the feedback non-directive? & 100.00\%\\
 & Does the feedback align with the feedback type?&88.33\%\\
\midrule
\textbf{Praise Feedback}   & Is the feedback appropriate for the sentence? & 100.00\%\\
                           & Is the praise specific to the details in the sentence? & 91.67\%\\
\bottomrule
\end{tabular}
\caption{Summary of Internal Audit Results for \system{}. All questions were evaluated on binary scales.}
\Description{Summary of internal audit results for Writor. The table lists three audit components: Goal Generation, Critique Feedback, and Praise Feedback, with evaluation criteria and success rates. Goal Generation achieved 100\% across all criteria, including relevance, specificity, and tailoring Goal 5 to instructor expectations. Critique Feedback showed high but varied rates: most criteria were at or near 100\% (e.g., topic fit, sentence appropriateness, non-directiveness), with slightly lower rates for goal alignment (96\%), usable text (96.67\%), inaccurate information (98.33\%), and feedback type alignment (88.33\%). Praise Feedback achieved 100\% for appropriateness and 91.67\% for sentence-specificity.}
\label{tab:audit_summary}
\end{table*}

We also compared our prompting approach against a single prompt for praise and critique, in which we asked the critique feedback to be non-directive and constructive. The results showed that our critique pipeline produces significantly longer (+46.9\%) and more specific (in terms of noun chunks) (+74.3\%) critiques compared to the baseline ($p < 0.01$).  \system{}’s critiques have slightly higher sentiment than the baseline (0.25 vs. 0.21), but the difference is not significant. Additionally, our approach generated praise that was significantly more specific (+20.2\%) and encouraging (+51.9\% sentiment). Details in Appendix~\ref{sec:baseline}.

\section{Expert Review of \system{}}
\label{sec:study2}
Our study aims to explore how writing center pedagogy translates and operationalizes into design guidelines (\S\ref{sec:designgoals}) for AI writing support tools. Our system, \system{}, is a functional prototype that operationalizes the design guidelines in practice. 

To evaluate this approach, we conducted an expert review with instructors, tutors, and AI researchers. We prioritized expert feedback for this stage of research because experts are uniquely positioned to assess the faithful translation of pedagogical guidelines (RQ1) into technical implementations and to validate \system{}'s pedagogical soundness (RQ2). Furthermore, this approach allows us to examine the role of writing center principles in guiding AI development and to identify where tools like \system{} might integrate into---rather than replace---existing writing support ecosystems (RQ3).

Our review was guided by the following research questions: 
\begin{itemize}
\item RQ1: In what ways does \system{} reflect or diverge from core principles of writing center pedagogy? \textit{(Translation of Design Guidelines)}
\item RQ2: How do participants rate the quality and usefulness of \system{}’s feedback? \textit{(Pedagogical Soundness)}
\item RQ3: What specific writing contexts or scenarios do participants view as appropriate for \system{} integration? \textit{(Potential for Integration)}
\end{itemize}

\subsection{Procedure}
\label{sec:study2procedure}
We conducted a mixed-methods study that had participants interact with \system{} and complete a structured survey evaluating \system{}'s feedback. We then conducted follow-up interviews with a subset of the participants. The study was conducted online, where participants accessed \system{} and the associated survey through a link.

\begin{figure*}[ht]
    \centering
    \includegraphics[width=1\linewidth]{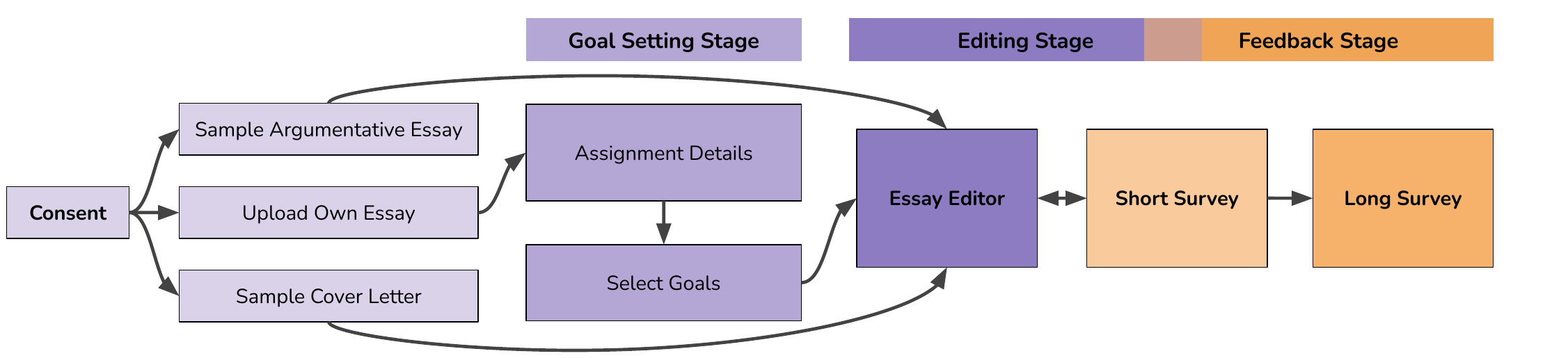}
    \caption{Study 2 Complete Workflow}
    \Description{Flowchart of Study 2 workflow with three stages: Goal Setting, Editing, and Feedback. Consent leads to three options: Sample Argumentative Essay, Upload Own Essay, or Sample Cover Letter. Upload Own Essay goes through Assignment Details and Select Goals before the Essay Editor; the other two go directly to the Essay Editor. After the editor, participants then complete a Short Survey followed by a Long Survey.}
    \label{fig:workflow}
\end{figure*}

After providing consent, participants chose between using \system{} with provided sample texts or their own writing samples. We provided two types of sample texts: 5 argumentative essays from The Michigan Corpus of Upper-Level Student Papers,\footnote{\url{https://elicorpora.info/main}} an open corpus of student writing samples, and 5 open-source cover letters from universities' career centers. Participants who selected a sample bypassed the goal-setting stage (\S\ref{sec:planning}) and proceeded directly to the editor, since setting artificial goals for texts they did not author was unnecessary. The goals for sampled essays were pre-selected by the researchers. However, those uploading their own samples completed the full workflow (Fig.~\ref{fig:workflow}).

After using \system{}, participants completed a structured survey evaluating \system{}'s feedback quality, features, and alignment with writing center pedagogy. We collected survey responses from two sources: a short survey directly in the editor interface and a longer survey after participants finished interacting with \system{}. Using two surveys allowed us to increase response rates and capture both immediate impressions and more reflective feedback. At the end of the long survey, participants could indicate their interest in participating in a raffle to win one of five \$25 Amazon gift cards. 

Participants could also indicate interest in a follow-up interview. Interested participants were contacted to schedule approximately 30-minute semi-structured interviews conducted via Zoom. They received a \$25 Amazon gift card within a week of completion. Interviews were audio-recorded, transcribed, and anonymized for analysis. This study was approved by the relevant Institutional Review Board.

\subsubsection{Data Collection}
We collected three types of data for analysis: (1) telemetry data, including feature usage, chat histories, and page navigation; (2) survey responses, from both the short and long surveys; and (3) interview transcripts from follow-up sessions with selected participants.

The short survey included one 5-point Likert-scale question on overall helpfulness, two open-ended questions about \system{}'s impact on writing education and its most and least useful features, one question on the participant's professional role, and an optional email field. The long survey consisted of 5 parts, categorized as: 

\begin{enumerate}
    \item \textbf{Feedback quality questions}, covering aspects including clarity, tone, usefulness, specificity, and willingness to adopt. We also include questions that asked participants to compare \system{}'s feedback with other feedback sources such as writing instructors and other generative AI tools.
    \item \textbf{Questions about the alignment of generated feedback to writing center principles}, grouped under the three core principles: writer-centered, process-oriented, and collaborative. 
    \item \textbf{Demographic and professional background}, including experience, institution type, and roles. 
\end{enumerate}

Both the short and long surveys can be found in Appendix \ref{sec:eval_survey}. All questions in the short survey were required except for leaving an email address. We did not require participants to answer all questions in the long survey.

For interviews, we contacted participants in the order they completed the long survey and provided their email addresses. We asked participants about their overall impressions of the tool and how it compared to other AI writing tools, how well its strategies align with their tutoring values, scenarios where it could be most helpful, suggestions for improvement, and any feedback type or goal they felt were missing. 

\begin{table}[ht]
\centering
\caption{Survey Participant Demographics. Note that participants were not required to answer demographics questions.}
\begin{tabular}{llc}
\toprule
\textbf{Characteristic} & \textbf{Category} & \textbf{Count} \\
\midrule
\multirow{6}{*}{Age} 
 & 18–24 years        & 3  \\
 & 25–34 years        & 7  \\
 & 35–44 years        & 9  \\
 & 45–54 years        & 3  \\
 & 55–64 years        & 1  \\
 & 65+ years          & 3  \\
\midrule
\multirow{3}{*}{Gender}
 & Women              & 19 \\
 & Men                &  6 \\
 & Non‑binary         &  1 \\
\midrule
\multirow{3}{*}{Race/Ethnicity}
 & White              &  9 \\
 & Asian              &  5 \\
 & Hispanic           &  2 \\
\midrule
\multirow{5}{*}{Experience}
 & Less than 1 year   &  7 \\
 & 1–5 years          &  5 \\
 & 6–10 years         &  5 \\
 & 11–15 years        &  3 \\
 & 15+ years          &  6 \\
\midrule
\multirow{3}{*}{Institution Type}
 & Public universities & 11 \\
 & Private universities&  7 \\
 & Industry            &  4 \\
\bottomrule
\end{tabular}
\label{tab:participant-characteristics}
\Description{Long survey participants’ demographic information, including age, gender, race/ethnicity, experience years, and institution type. Participants were mostly aged 25–44, majority women, predominantly White and Asian, with varied experience levels and varied institution types (private universities, public universities, or industry).}
\end{table}

\subsection{Participants}

We recruited participants through multiple channels to reach writing educators and AI researchers. We sent direct email invitations to writing center administrators at U.S. universities, asking them to share the study information with their staff. We also distributed recruitment flyers at academic conferences related to writing studies and composition, posted recruitment materials on social media platforms and professional networks, and leveraged personal and professional connections within the writing studies and educational technology communities. Additionally, we employed snowball sampling, encouraging participants to share the study with qualified colleagues. We closed the survey when there were no new responses for a week. 

A total of 44 eligible participants completed the study between Jun 10, 2025 and July 28, 2025. To be eligible, participants needed to be between 18 and 80 years old, fluent in English, and have greater than 3 months experience tutoring or teaching writing and/or building AI-writing tools. Given the survey was distributed widely, we applied several exclusion criteria to ensure data quality and authenticity:

\begin{itemize}
\item Responses reporting conflicting professional roles or expertise across multiple role-related questions (e.g., inconsistent answers about being a tutor, instructor, or researcher) were excluded;
\item Responses utilizing atypical email domains\footnote{E.g., domains similar to \texttt{hahahahahaha.io} (not the actual domain)} were excluded;
\item Multiple submissions containing substantially similar or identical open-ended and close-ended responses across different email addresses were excluded;
\item Submissions exhibiting identical ratings (e.g., answering 'not at all' to every question) across all quantitative survey items were excluded;
\item Responses with blank open-ended fields or placeholder content (e.g., ``.'', ``NA'') in all required responses were excluded.
\end{itemize}

After applying exclusion criteria, survey responses from 30 participants were included in the final analysis. Of these, 26 completed both the short and long surveys. 23 participants identified as writing instructors, 11 as writing tutors, and 6 as having AI writing research experience (see Table ~\ref{tab:participant-characteristics} for more demographic information and Appendix ~\ref{sec:append_results} for a detailed role breakdown). Participants could select multiple roles. Participants engaged with \system{} using different text types: 8 used a provided argumentative essay, 14 used a provided cover letter, and 8 uploaded their own writing. 11 participants were selected for follow-up interviews. Their backgrounds can be found in Table ~\ref{tab:interview-background}. Five of the 11 interviewees also participated in Study 1 (\S~\ref{sec:formative}).

\begin{table*}[ht]
\centering
\caption{Interview Participant Background}
\begin{tabular}{llll}
\toprule
\textbf{Participant} & \textbf{Role}                   & \textbf{Institution}           & \textbf{YOE} \\
\midrule
P1  & Tutor                       & Public University              & 5     \\
P2  & Tutor                       & Private University             & 2     \\
P3  & Tutor                       & Private University             & 3     \\
P4  & Tutor                       & Private University             & 1     \\
P5  & Instructor                  & Public University              & 5     \\
P6  & Tutor                       & Public University; Industry    & 5     \\
P7  & AI Researcher                       & Industry                       & 2     \\
P8  & Instructor                  & Public University              & 10+   \\
P9  & Tutor; Instructor; AI Researcher                                 & Private University             & 10+   \\
P10 & Instructor; AI Researcher              & Public University              & 15+   \\
P11 & Instructor                  & Private University             & 15+   \\
\bottomrule
\end{tabular}
\label{tab:interview-background}
\Description{Interview participant background, including their roles, institution types, and year of experiences.}
\end{table*}

\subsection{Analysis}

\subsubsection{Quantitative Analysis}

We conducted descriptive and comparative analyses using participants' responses to the structured survey (Appendix \ref{sec:eval_survey}). We grouped responses by self-identified professional roles (tutor, writing instructor, AI researcher) and by attitude toward AI in writing. Because participants could select multiple roles, some analyses included overlapping categories, meaning individual responses may appear in more than one group. To assess pedagogical alignment, we analyzed aggregated responses across three core dimensions (writer-centered, process-oriented, collaborative) and examined sub-component ratings and variability using standard deviations. We elected to not perform or report inferential statistics (e.g., paired \textit{t}-tests) on our data for two reasons. The first was that we recruited from multiple expert populations and therefore had a relatively small sample size for each group (e.g., 6 AI researchers) and would be performing tests without the requisite power. Second, our study did not focus on comparisons with existing systems but rather aimed to assess experts' reactions to non-directive intelligent writing support. Because existing systems provide qualitatively different support, we did not compare \system{}'s feedback against a baseline or control group and therefore did not have a null-hypothesis to reject or accept based on statistical significance. When reporting quantitative findings, we include insights from our interviews to deepen our overall review of \system{}.

\subsubsection{Qualitative Analysis}

We analyzed open-ended survey responses and interview transcripts using a similar procedure as \S\ref{sec:interview_procedure}. One researcher performed initial open coding on all transcripts, identifying recurring ideas and themes related to feedback quality, pedagogical alignment, system usability, and areas for improvement. To generate possible new codes and assess consistency with initial codes, a second researcher independently coded one interview and reviewed initial codes with the first researcher. Through weekly collaborative discussions over the course of a month, we refined and consolidated the code structure. Both researchers iteratively developed a shared codebook, resolving discrepancies through consensus. Once finalized, the primary researcher applied the codebook across all transcripts to ensure consistency.

\paragraph{Positionality} 

The background and experiences of our research team impacted our data collection and interpretation. Our team consisted of two former writing center tutors who are now AI/HCI researchers along with researchers in English studies and HCI. Our network gave us access to participants with diverse experiences and backgrounds (writing tutors, writing instructors, and AI researchers). Some participants with anti-AI attitudes revealed that they still chose to participate because of the team's connection to writing studies and the trust they placed in them. In addition, the first author's background in writing tutoring allowed them to quickly establish rapport and connect with discussions of writing instruction during interviews. In coding the interviews, the team's background allowed us to interpret participants' responses from multiple perspectives: our tutoring experience enabled understanding of pedagogical terminology, theoretical frameworks, and the practical challenges of using systems like \system{}, while our AI/HCI expertise allowed us to contextualize technical critiques regarding feedback specificity, algorithmic reasoning, and interface design.

\section{Expert Review Findings}
% keep an eye on which quote is not labeled with participant number

\subsection{RQ1: In what ways does \system{} reflect or diverge from core principles of writing center pedagogy?}
\label{sec:RQ1_findings}

\begin{figure*}[ht]
    \centering
    \includegraphics[width=1\linewidth]{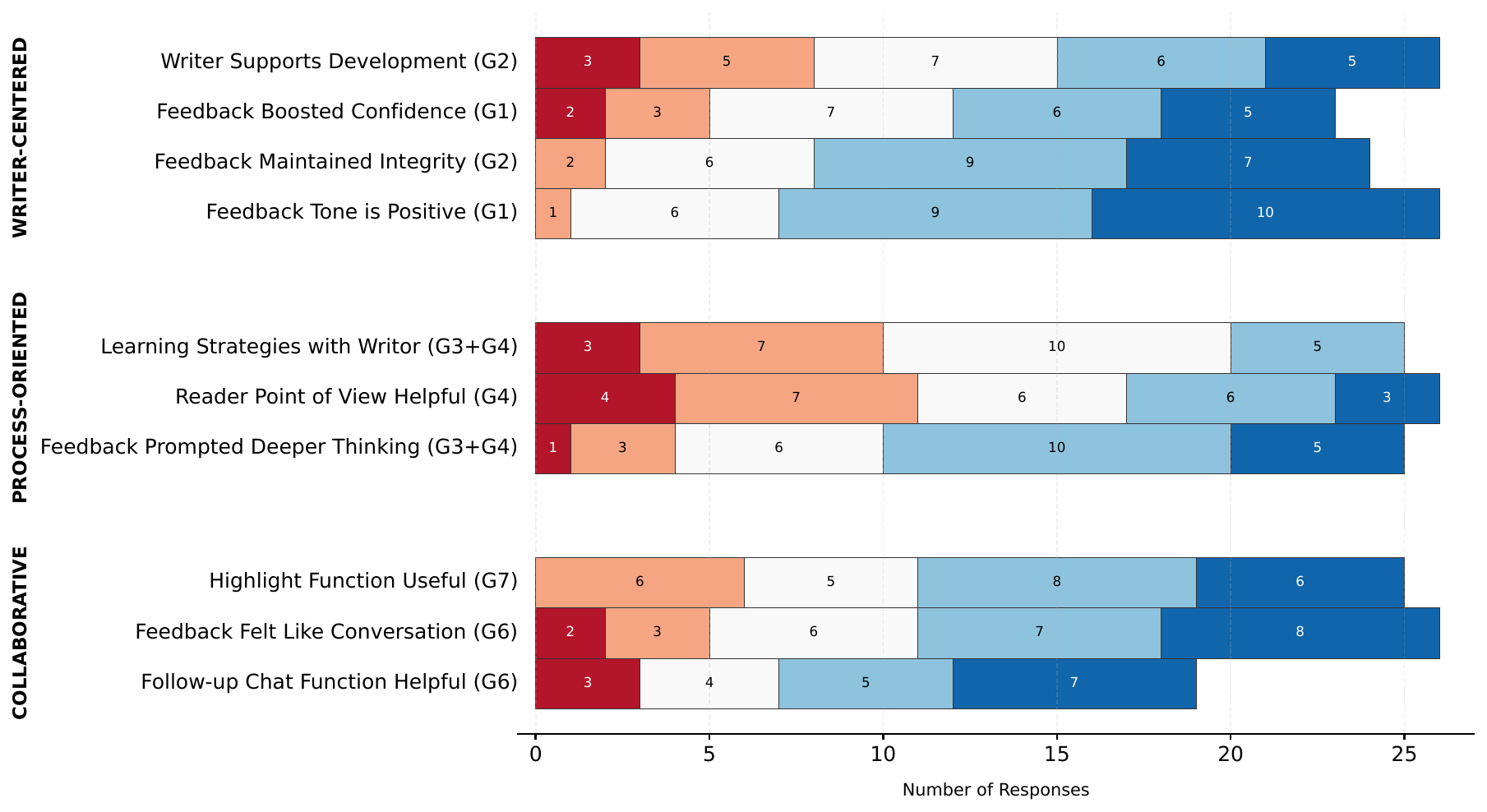}
    \caption{Participant ratings of \system{}'s alignment with writing center pedagogical principles across three key dimensions: writer-centered, process-oriented, and collaborative. Color indicates their agreement levels, and the number inside each segment indicates the number of participants. Below each statement is the associated design guideline. As a note, G5 (Aligns with Goals) was not presented in the survey because participants were not required to go through the goal-setting stage. We also include supplementary analyses by participants’ writing-instruction backgrounds (Fig.~\ref{fig:wc_role}) and by their attitudes toward AI (Fig.~\ref{fig:wc_attitude}) in Appendix~\ref{sec:append_results}.} 
    \Description{Stacked horizontal bar chart of feedback quality ratings grouped into Writer-Centered, Process-Oriented, and Collaborative. Writer-Centered (low to high): Writor Supports Development, Boosted Confidence, Maintained Integrity, and Tone Positive. Process-Oriented: Learning Strategies, Reader POV Helpful, and Feedback Prompted Deeper Thinking. Collaborative: Highlight Function Useful, Feedback Felt Like Conversation, and Follow-up Chat Helpful.}
    \label{fig:pedagogy_alignment}
\end{figure*}

Fig. ~\ref{fig:pedagogy_alignment} reports overall ratings for each dimension associated with writing center pedagogy and design guidelines. Interview responses generally revealed strong recognition of writing center pedagogical approaches in \system{}. In the interview, nine participants noted how \system{} is similar to writing centers. P3 observed: \textit{``what they[\system{}] said is like very standardized writing center language... these strategies are like, good, writing center strategies.''} P5 confirmed: \textit{``the questions raised by... the system... those questions are pretty much standard like those are some of the things that we'll typically ask.''} P6 noted the authentic sound: \textit{``it repeats the kinds of language that we use in the writing center.''} Moreover, P9 memorably described \system{} as feeling \textit{``like a writing center at 3 AM.''}. Below we go into detailed responses for each dimension. 

\subsubsection{Writer-centered}
\label{sec:writerCenteredS2}

As shown in Fig. ~\ref{fig:pedagogy_alignment}, ``Feedback tone is positive'' and ``Feedback boosted confidence'' both had high agreement ratings, demonstrating that \system{} instantiates \textbf{G1} (Fig. ~\ref{fig:guidline}). In the interview, seven participants strongly valued \system{}'s positive, confidence-building approach. P3 noted: \textit{``it [the feedback] seems to be very encouraging... [and] helps the writer feel confident in writing, which I think is probably one of the most important things.''} \system{}'s balanced approach---combining both praise and critique---distinguished it from other AI tools, as P1 appreciated: \textit{``The other thing that's annoying about using ChatGPT...[is that it] always give me something to change, even when it really doesn't need to be changed anymore.''} P11 valued the potential to reinforce good practices through praise: \textit{``[it's] not just saying like great job, but pointing out, what about it that is good. So that students kind of reinforce whatever skill or principle that's being applied there.''} 

Participants agreed that ``feedback maintained integrity'' of their voices and ideas, echoing \textbf{G2}. Beyond the 5-point Likert scale questions, 84\% (21/25) of participants noted that \system{} never generated text they could copy and paste. After manual review of the chat logs for the four participants who reported getting usable text, we saw that in five of the 94 total conversation turns\footnote{A conversation turn includes a user's prompt and an agent's response.} among these participants, \system{} generated copyable text, either a full or partial sentence (such as how to start a sentence). While in some longer conversations \system{} eventually provided text, participants generally saw that \system{}'s value was in discouraging shortcuts. As P8 described it: \textit{``if there's a platform available to discourage shortcuts and encourage that rhetorical, critical thinking while still taking advantage of the fact that students are using generative AI, then that is a value.''} In the survey responses, three participants believed that non-directive approaches were \textit{``what AI for writing needs to be as it continues growing.''}

\subsubsection{Process-Oriented}
\label{sec:processOrientedS2}

In the interviews, participants recognized \system{}'s effective use of questioning to promote deeper thinking. While the feature was designed to sustain conversation (\textbf{G6}), participants valued the questions for supporting process-oriented revision. P4 noted: \textit{``it sort of prompts the students to think about different ways they can write something.''} In the submitted surveys, 5 participants explicitly praised the prompting questions, highlighting their usefulness in guiding reflection and encouraging thinking in the revision process. Reflecting these responses, ``Feedback prompted deeper thinking'' had high agreement among survey respondents in the process-oriented category.

In addition to the features at the editing stage, an important process-oriented design choice was the goal-setting stage (\textbf{G5}). In the interviews, two participants appreciated \system{}'s goal-setting functionality for helping writers understand their writing objectives. P7 noted: \textit{``I did really like the goal setting part, because the reality is that if you are not writing an expository piece of writing, you actually probably have a weaker sense of what the goals are,''} and stated that the goal setting part is \textit{``the most helpful''} in their experience using the tool. However, three participants noted that students might be unfamiliar with the concept of setting writing goals or uncertain about how to articulate them, often having a vague idea without knowing exactly what they want to focus on.

\begin{figure*}[ht]
    \centering
    \includegraphics[width=0.63\linewidth]{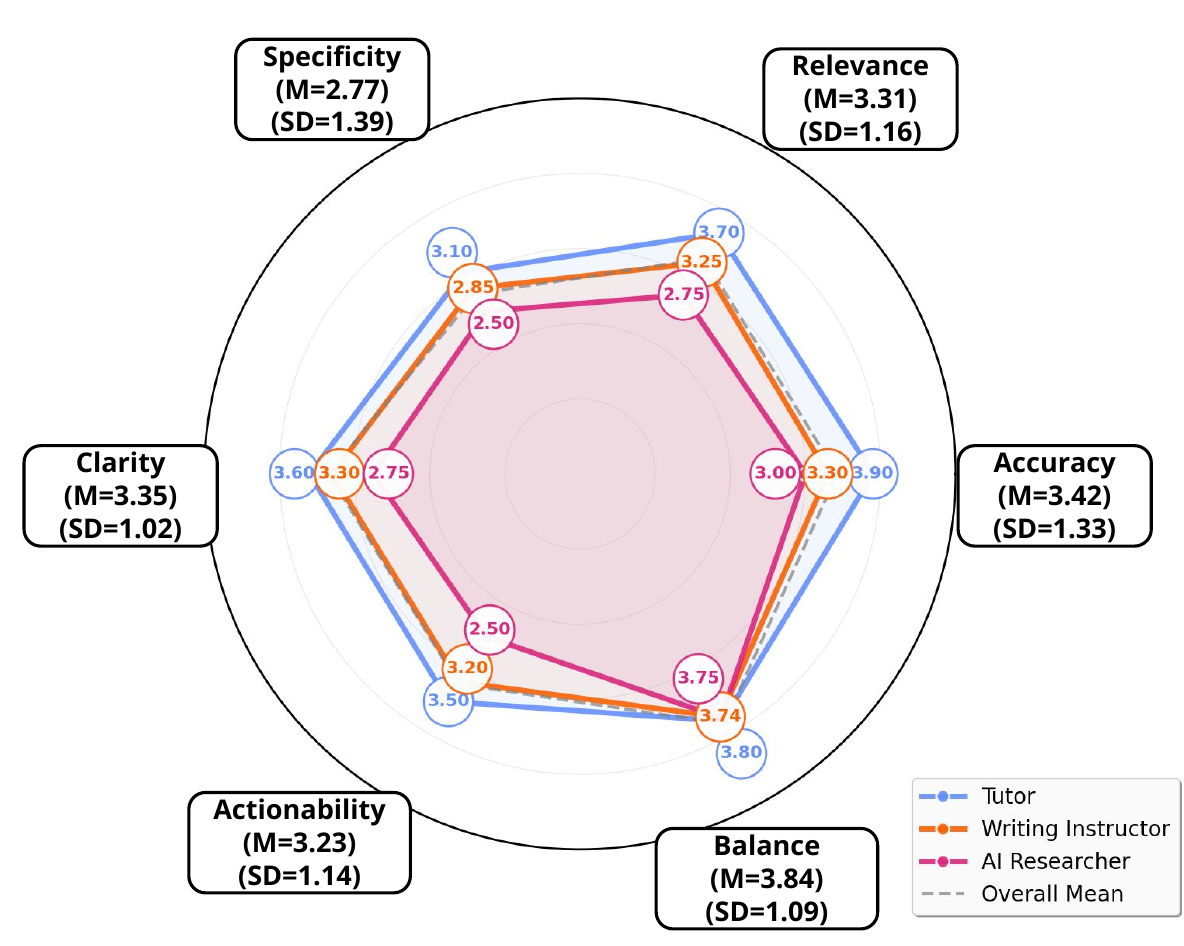}
    \caption{Mean feedback quality ratings across six dimensions by expert type. Balance is rated the highest, and Specificity the lowest. We also include supplementary analyses by participants’ writing-instruction backgrounds (Fig.~\ref{fig:quality_role}) and by their attitudes toward AI (Fig.~\ref{fig:quality_attitude}) in Appendix~\ref{sec:append_results}.}
    \Description{Radar chart showing mean feedback quality ratings across six dimensions by expert types. The dimensions are: Specificity, Relevance, Accuracy, Balance, Actionability, and Clarity; expert types are: Tutor, Writing Instructor, and AI Researcher. Balance has the highest overall rating (M=3.84), followed by Accuracy (M=3.42), Clarity (M=3.35), Relevance (M=3.31), Actionability (M=3.23), and Specificity (M=2.77). Tutors mostly rate highest across most dimensions, while AI Researchers rate lowest on several, especially Specificity and Actionability. }
    \label{fig:feedback_quality}
\end{figure*}

\subsubsection{Collaborative}
\label{sec:collabrativeS2}

Participants generally supported the follow-up chat feature and felt that interacting with \system{} was conversational. Overall, 22 of the 30 participants engaged with the chat function. Their average engagement time was 6.38 minutes ($Median = 2.5, SD = 10.78$). On average, they completed 10.91 conversational turns with \system{} ($SD = 8.77$). Three participants emphasized the importance of the chat function. P8 observed how the chat functionality promotes engagement, aligning with G6: \textit{``you've forced students to engage with the with the chat... Because the student has to engage... I think that students would use that opportunity to chat more... instead of just taking a shortcut which doesn't teach them anything.''}

\subsection{RQ2: How do participants evaluate the quality and usefulness of \system{}’s feedback?}

\subsubsection{Feedback Shows Strong Balance and Positive Tone}

Participants consistently praised \system{} for delivering a well-balanced mix of praise and constructive critique in a positive tone. While related to participants’ appreciation of positivity and confidence-building in RQ1's findings (\S\ref{sec:RQ1_findings}), this subsection highlights how balance and tone were assessed as core quality dimensions of \system{}’s feedback.

Seven interviewees emphasized the role of praises: P8 noted it \textit{``gave the student confidence''} and P4 highlighted that \textit{``the compliment section\ldots{}is really important in writing.''} These impressions were mirrored in quantitative ratings (Fig. ~\ref{fig:feedback_quality}). Balance---the mix of positive and constructive comments---earned the highest score overall ($M = 3.84, SD = 1.09$, with 1 = Not at all helpful and 5 = Extremely helpful), with all professional groups rating it favorably. In survey responses, nine participants identified positive tone in the constructive critiques as \system{}’s greatest strength.

Interestingly, participants rated the overall helpfulness of \system{}’s feedback as only somewhat helpful (mean and median = 3.00, SD = 1.02). While this overall rating was moderate, participant comments pointed to specific factors that may have influenced their assessments and room for future improvements or customization.

One factor was balancing specificity in generated feedback. In the interviews, while two participants found certain comments too general (P2: \textit{``This feedback does feel a little, general''}), three felt they were overly specific, as P2 noted \textit{``I would feel pressured to do that [changes] because it highlighted that...[in a] such specific feedback.''} This range of reactions suggests that mismatched levels of detail may have lowered overall ratings and points to the need for calibrating feedback to individual writer needs.

\begin{figure*}[ht]
    \centering
    \begin{subfigure}[t]{1\linewidth}
        \centering
        \includegraphics[width=\linewidth]{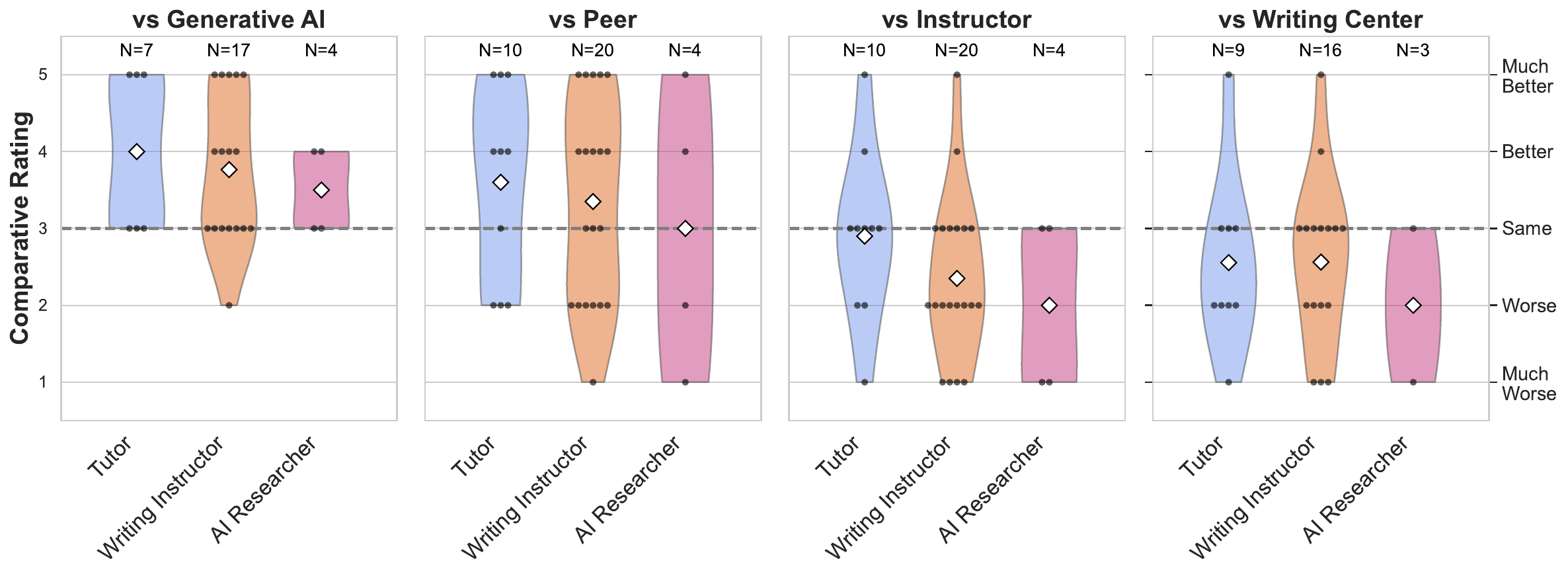}
        \caption{Grouped by expert role}
        \label{subfig:compare_sources_roles}
    \end{subfigure}
    \vskip 1em
    \begin{subfigure}[t]{1\linewidth}
        \centering
        \includegraphics[width=\linewidth]{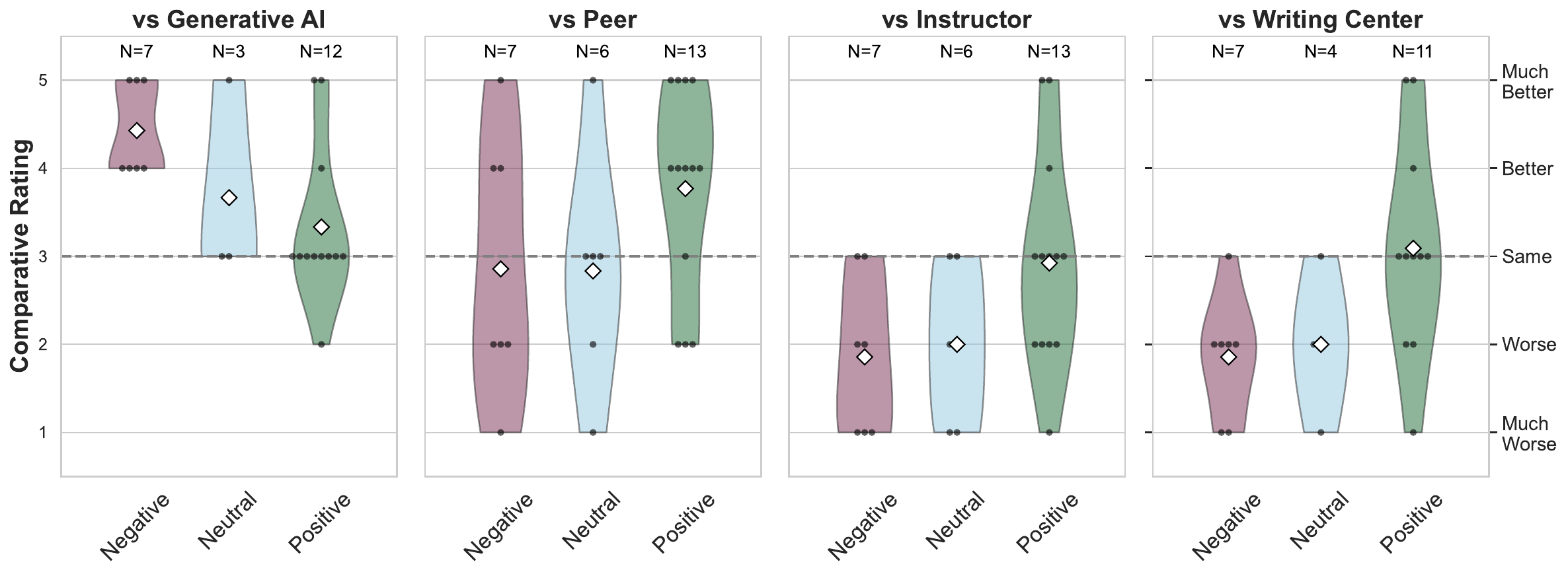}
        \caption{Grouped by participant attitude toward AI in Writing}
        \label{subfig:compare_sources_attitudes}
    \end{subfigure}
    \caption{Perceived quality of \system{}’s feedback compared to instructor, peer, writing center, and generative AI feedback. Dashed line represents the ``same'' rating, diamond shape represents the mean value. Participants generally found \system{} better than other generative AI tools, yet worse than writing center or instructor's feedback. Note that these are perceptual ratings and participants were only given text examples of generative AI (e.g., ChatGPT, Claude, etc.).} 
    \Description{Two sets of violin plots showing perceived quality of Writor’s feedback compared to generative AI, peer, instructor, and writing center feedback. Panel A groups results by expert role---Tutor, Writing Instructor, and AI Researcher. Tutors generally rate Writor higher than AI Researchers. All but one comparison shows ratings above ‘same’ when compared to generative AI, with group means near the ‘better’ line; ratings are lower when compared to instructors or writing centers. Panel B groups results by participants’ attitude toward AI in writing---Negative, Neutral, Positive. Positive-attitude participants tend to give higher ratings, especially when comparing Writor to human writing support. Participants with negative AI attitudes rate Writor between ‘better’ and ‘much better’ compared to generative AI, higher than both neutral and positive groups.}
    \label{fig:compare_sources}
\end{figure*}

\subsubsection{\system{} is preferred over other AI and might complement human feedback}

Although participants with negative attitudes toward AI often rated \system{} lower on alignment and feedback quality (see Appendix~\ref{sec:append_results}), \textbf{\system{} was perceived as preferred over other AI tools (particularly for participants with a negative attitude towards AI) but fell short of human feedback sources, suggesting its optimal role as a complement to---rather than a replacement for---human tutoring.} In interviews, seven participants expressed strong preferences of \system{} over other generative AI tools, while citing concerns about existing generative tools replacing human thinking. P3 explained that \textit{``a big reason why a lot of tutors don't like generative AI is that they don't want the AI to take away the human thought in the writing process.''} In contrast, participants valued that \system{} preserved student agency and original text due to its non-directive nature. P10 appreciated that \textit{``this is an invitation for students to analyze their writing, and it seems like the goal is not to make the human or the thinker obsolete, but to give them more information about what their writing is doing.''} The non-intrusive approach was particularly valued by P10: \textit{``I like that\ldots{}whatever interaction is happening is off to the side that the writing itself isn't being touched.''} In the survey responses, 8 participants expressed positive attitudes toward systems like \system{} specifically because they avoid direct text generation.

While \system{} rated lower than instructors and writing center tutors in direct comparisons (Fig. \ref{subfig:compare_sources_roles}), interviews framed this feedback as emphasizing the complementary nature of AI support rather than a limitation. P1 recognized the distinct value humans provide: \textit{``I don't think it [the system] totally does the same thing that we do in person\ldots{}...[we do] like moral support and like encouragement.''} P9 similarly highlighted human tutors' unique capacity for empathy: \textit{``it's easier to deal with as a person, because I can see that they're like frustrated, whereas I think, like a agent would not be able to.''} Below, we discuss the opportunities participants saw for \system{} to complement existing human writing support.

\subsection{RQ3: What specific writing contexts or scenarios do experts view as most appropriate for \system{}'s integration?}
\label{sec:RQ3}

Participants identified four primary contexts where \system{} could be integrated: before writing sessions, modeling peer reviews, self-directed learning, and after writing sessions. Notably, participants' focuses slightly differed based on professional background (e.g., writing tutors mostly focused on \system{} support within writing centers). 

\paragraph{Before Writing Sessions} The most frequently cited context was using \system{} before tutoring sessions, with 6 participants (including 5 writing tutors) citing this use case. P1 suggested pre-session use could improve efficiency: \textit{``if they entered some information in on a form, and then, like ran it through this thing before the appointment that might actually get them moving a little better.''} This reflects the view that students could arrive more prepared and already familiar with writing center strategies and session formats. P5 highlighted the comparative value: \textit{``students can use this tool and then come into the writing center, and then compare the feedback\ldots{}they can also analyze like, hey, how is this conversation with the [system] different from the conversation that I have with my tutor.''} This comparative value highlights how \system{} and human tutors together can broaden the role of writing support by encouraging students to reflect on how different feedback sources complement one another. Ultimately, participants saw \system{} as an additional aid but not a replacement for writing tutors: supporting rather than supplanting human feedback.

\paragraph{Modeling Peer Reviews} 3 participants identified \system{}'s potential to model peer review practices for student writers. P6 explained: \textit{``it could be useful in modeling what peer review could look like\ldots{}the only model that my students would have for this kind of feedback is for me, and I think they might take it differently when it comes from an instructor.''} P10 also noted: \textit{``students aren't sure how peer review is helpful for them as writers. So I would think that this kind of parsing and the interaction might be helpful.''} Together, these comments suggest that \system{} may help students better understand the purpose and process of peer review by providing a practice model distinct from instructor feedback that can be distorted by power dynamics.

\paragraph{Self-Directed Learning} \system{} was viewed by 3 participants as appropriate for independent practice. P1 compared it to personal writing tools: \textit{``this is very similar to how how I use ChatGPT... I could see this as like almost like a regular way, a regular thing you use while you're writing.''} P9 identified broader applications and improving accessibility: \textit{``this would be a game changer for people who are not confident about their writing skills even after graduation, or having never gone to college.''}

\paragraph{After Writing Sessions} 2 participants valued \system{} for extending learning beyond session constraints. P2 explained: \textit{``it would make more sense to use it after a session, because often we'll have a session, and, like our sessions were always 45 min\ldots{}And then\ldots{}they're like, Oh, it's over.''} \system{} could provide \textit{``more time to kind of look at every single potential question that the [student] has without taking the writing tutor[s' time]\ldots{}there's not enough time for every single question.''}

% DONE final checking 

\section{Discussion}

Writing centers have long recognized that giving students ready-made sentences or paragraphs, while seemingly helpful, ultimately undermines the learning process of writing by reducing writer agency and ownership \cite{2012oxford}. Practically, too, ready-made writing reduces practicing writing. Building on this pedagogical approach, we propose design insights to a form of intelligent writing support that focuses on non-directive feedback buttressed by writing center strategies. 

\subsection{Designing for Writing Process}

The default text-generation capabilities of LLMs create a tension in how to scaffold writing process when systems can rapidly produce ready-to-use text. While previous literature identified risks of AI-generated text, including plagiarism \cite{rainie_watson_2025, santiago_utilization_2023, malik_exploring_2023, semrl_ai_2023, cotton_chatting_2023, almaiah_measuring_2022}, reduced writer agency \cite{katy_creative, reza2025cowritingaihumanterms, Stark2023, 10.1145/3698061.3726907}, and discouraged critical thinking and reflection \cite{marzuki_impact_2023}---\system{} shows that AI can provide meaningful support without directly producing text for writers, thereby mitigating some of the aforementioned concerns. This design stance situates \system{} within recent calls to develop generative AI to support human cognition rather than supplant it \cite{tankelevitch_tools_2025, pokkakillath_chatgpt_2023, siddiqui_ai_2025, arnold2021generative}.

This work contributes to HCI literature on human-AI co-writing by creating a design framework arising from established writing center pedagogy to scaffold writing processes. Prior HCI work has often framed AI as a collaborator that shares the labor of text production \cite{wordcraft_yuan, mirowski_co-writing_2023, yeh_ghostwriter_2025, choe_supporting_2024}, whereby the system and user take turns generating content. In contrast, we utilize the idea of generating non-directive feedback to facilitate writers' own production of text. Other systems have moved toward scaffolding by offering inspiration for metaphors \cite{Gero19metaphoria} or scientific writing ideas \cite{sparkkaty}, yet these still rely on providing content for users to select. Closer to our approach are systems that employ Socratic questioning \cite{kim2023repurposingtextgeneratingaithoughtprovoking, arnold2021generative}, continuous summaries \cite{10.1145/3526113.3545672}, or use examples \cite{choe_supporting_2024} to provoke reflection. In our work, we translate writing center principles into innovative strategies that inculcate reflection beyond these approaches. These include: (a) using reader-perspective feedback, examples, and analogies to guide revision without supplying content (process-oriented); (b) pairing praise with emphatic critiques oriented towards writer's goals (writer-centered); (c) inviting open-ended, goal-aligned dialogue to keep the system accountable to what the writer wants to work on (collaborative).

Broadly, this work highlights how pedagogy can be translated into design guidelines that provide a foundation for designing interactive systems like \system{}. While prior systems often incorporate individual design tenets---such as non-directive feedback through questioning and examples \cite{kim2023repurposingtextgeneratingaithoughtprovoking, arnold2021generative, choe_supporting_2024}, empathy \cite{empathy_learning}, conversations \cite{kim-etal-2025-voice, talaei_storysage_2025}, reader perspective \cite{writer_persona}---these implementations are typically piecemeal. In contrast, writing center pedagogy offers an integrated framework that inherently encodes boundaries around agency and levels of assistance. Translating such framework into system design allows us to move beyond isolated features to a consistent and proven theory of support.

\subsection{Design Implications From Expert Review}

\subsubsection{Designing for Trust Among AI-Skeptical Educators through Non-Directivity}

Our expert review revealed a notable pattern: participants with the most negative attitudes toward AI consistently perceived \system{} as better or much better than standard generative tools.\footnote{In the survey, the examples provided for standard generative AI tools were ChatGPT, Claude, Gemini, and DeepSeek.} As P3 in the expert evaluation explained, tutors ``don’t want the AI to take away the human thought in the writing process.'' Participants saw \system{} as avoiding this pitfall. They emphasized that \system{}'s non-directive design preserved student agency, kept the writer’s original text intact, and prompted reflection rather than replacement. These reactions suggest that skepticism toward AI in writing education is not entirely a rejection of technology itself, but of modes of AI intervention that diminish agency, originality, or critical thinking.

This insight points to a broader design implication: AI writing tools can earn trust among AI-skeptical educators by making students' agency visible and protected \cite{han_teachers_2024, lyu_understanding_2025}. More broadly, grounding system design in an explicit pedagogical framework may itself contribute to trust, signaling to educators that the tool is built to reinforce, rather than erode, the learning process. This non-directive design philosophy can extend beyond writing instruction to other domains susceptible to AI automation to build trust among educators, such as computer science education and mathematics. For instance, rather than generating functional code blocks or solving equations directly, an agent might be restricted to providing debugging logic (if the goal is not to learn debugging), while a math assistant could focus on identifying conceptual misconceptions. In these contexts, as in writing, trust is established not by the AI's capability to perform the task, but by its distinct refusal to do so in favor of guiding the students' own process in learning.

Nevertheless, careful attention is still needed when relying on non-directive design. Even when a model refrains from generating text, it still shapes the writing process by deciding what to critique, what to praise, and what to disregard. By subtly directing revision, agency can be shaped rather than supported \cite{carroll_characterizing_2023}. Although \system{} uses goal-setting to help ensure feedback reflects writers’ stated intentions---an approach aligned with ethical nudging \cite{sanchez_chamorro_ethical_2023}---the inherent opacity of LLMs complicates how faithfully this alignment is enacted and limits the transparency necessary for users to understand or evaluate those guiding choices. As we integrate these tools into educational settings, systems should help students be aware of these limitations. Given these inherent constraints of AI-only feedback, co-support between human instructors and AI agents presents a promising direction for addressing these tensions, which we examine next.

\subsubsection{Designing AI Tools within Human Writing Ecosystems}
\label{sec:augment_discussion}

\system{} was perceived as a complement to, rather than a replacement for, human writing instruction (\S\ref{sec:RQ3}). While AI systems like \system{} can offer scalable feedback and will potentially improve as the underlying technology continues to mature, human tutoring remains irreplaceable for fostering interpersonal relationships, empathy, and individualized support. Thus, the vision we advance is not substitution but augmentation, that AI writing support tools should extend and complement human writing support. 

Participants identified four potential integration contexts for \system{}, with two directly supporting traditional tutoring sessions: to use for preparation before sessions and follow-up after sessions. These integration contexts open the door to specific technical enhancements that could strengthen \system{}, increasing the accessibility and continuity of writing support. For pre-session preparation, \system{} can add an export function that allows students to bring edit logs and interaction histories to tutoring appointments. Tutors also benefit from seeing a record of the student’s prior interactions with \system{}, giving early insight into the student’s needs as well as easing the introduction of writing center pedagogy. For post-session editing, \system{} can be redesigned to support continuity with past tutoring conversations, extending learning beyond the time constraints of writing sessions and into hours when the writing center is inaccessible. Rather than \system{} automatically highlighting sentences for feedback, students and tutors could together identify specific areas to continue working on beyond the tutoring session. These improvements would transform \system{} from a standalone tool into a bridge between human and AI writing support.

In addition to direct integration with writing centers, participants also identified that \system{} could be use to model peer reviews for student writers. Since participants recognized \system{}'s alignment with writing center principles and non-directive feedback approaches, students could learn effective feedback practices through interaction with the system. While \system{} was not designed as a trainer for giving feedback, its demonstration of constructive, process-oriented commentary could serve as implicit modeling for how students might approach reviewing their peers' work.

\subsubsection{Integrating Specific Praise into Feedback Design}
\label{sec:positivity_discussion}

Writing is inherently vulnerable work, as writers expose their thinking and ideas through their texts, and productive feedback must create a supportive environment for growth \cite{2012oxford}. The importance of praise emerged as one of \system{}'s most valued features. Prior work showed that giving students praise enhances their motivation \cite{campean_examining_2024}, engagement \cite{campean_examining_2024}, acceptance of feedback \cite{patchan_validation_2009}, and improvements in grades \cite{faulconer_impact_2022}. In our study, participants rated ``balance''---defined as the appropriate mix of positive (praise) and critical feedback---as the highest dimension across all professional groups, with interview participants consistently praising the system's encouraging tone. The effectiveness of praise extends beyond mere encouragement to active learning support. In the interview, participants believed that positivity helps to reinforce positive behaviors. 

At the same time, praise may risk collapsing systems into sycophancy. LLMs have a tendency to generate excessively agreeable responses to align with users' expectations \cite{sharma_towards_2025}. In the writing context, this might look like overly generous praise. To avoid sycophancy, we designed \system{}'s feedback to make praise specific and integrated with critiques.  For instance, positive comments highlight particular achievements (e.g., ``your last sentence clearly summarizes the paragraph and strengthens your argument'') rather than generic approval (e.g., ``great job'') inspired by writing center pedagogy \cite{2012oxford}, and are deliberately balanced with a slightly greater number of constructive suggestions. Our audit (\S\ref{sec:audit}) further demonstrating that \system{}'s praises are more specific than a generic prompt for positive feedback.

We argue that integrating praises into AI writing feedback is not only about making systems ``nicer'' but about supporting writers' development. We suggest future feedback systems to prioritize specific positivity as a core design principle (e.g., praise for particular achievements, such as a well-structured thesis) .

\subsubsection{Designing for Adaptive Granularity in Feedback}

Specificity was the lowest-rated dimension in \system{}'s feedback. In qualitative findings, some participants found comments too general to be actionable, while others felt ``pressured'' by feedback that was so specific it bordered on being directive. This discrepancy suggests that a static approach to feedback generation may be insufficient for the diverse needs of writers, suggesting a need for adaptive specificity for feedback-giving systems.

To address this, future systems should move beyond single-turn prompting and implement user-controlled granularity for feedback specificity. Interfaces could offer mechanisms, such as a slider, allowing writers to explicitly request the level of detail they need. While highly specific feedback carries the risk of becoming directive, systems can mitigate this by utilizing process-oriented strategies, such as providing analogies and examples (\textbf{G3}) or using reader-perspective style feedback (\textbf{G4}), to ensure that the feedback remains a scaffold for user reflection rather than a correction to be passively accepted.

\subsection{Alternate Design of Writor}
\label{sec:alternativedesign_discussion}

\system{} represents one possible implementation of our design guidelines for integrating writing center pedagogy into intelligent writing support systems. However, the guidelines we developed (Fig. ~\ref{fig:guidline}) could inform alternative designs or modalities. For example, rather than providing structured feedback cards, a system could represent feedback exclusively through dialogue (e.g., a conversational agent). This approach would more directly mirror the conversational nature of writing center sessions, allowing for immediate clarification via follow-up questions, thereby suiting  a writer's specific needs. Extending the idea of dialogue, a system could be entirely voice-based \citep{kim-etal-2025-voice} to  mirror person-to-person tutoring conversations. 

These guidelines can also inform the design of tools that integrate more directly with existing writing center sessions. Rather than standalone systems like \system{}, alternative designs could enhance and scaffold writing center sessions. We have pointed to potential new features to support before and after writing sessions in \S\ref{sec:augment_discussion}, as suggested by the participants. However, we imagine that tools that work during the tutoring sessions would also be helpful. For instance, an AI transcription and summarization tool could capture session dialogue in real time, showing whether goals were met and highlighting key discussion and feedback areas. Tutors and students could then revisit these summaries to reflect on the process and decide on next steps. Aligned with our guidelines, future tools could use empathic language to highlight strengths while focusing on higher-order concerns.

\subsection{Limitations and Future Work}

Here we discuss four limitations and suggested directions for future research.

First, our study focused exclusively on expert viewpoints. Because evaluating whether \system{} is designed to support effective writing processes is fundamentally a question of pedagogical fidelity, this stage of research relied on experts rather than students. While this approach provided valuable insights into understanding pedagogical practices and evaluating \system{}'s effectiveness, alignment, and potential for integration, it may not capture how students actually experience writing centers and \system{}. For example, students without any writing center experience may find the feedback confusing, as the feedback approach might be unusual for them. Future work could conduct interviews with student writers who have used writing centers and comprehensive user studies with them to understand real-world usability and learning outcomes of systems such as \system{}. Additionally, a longitudinal field deployment could investigate whether students internalize the revision strategies modeled by \system{} and apply them to subsequent writing tasks without AI support.

Second, participants rated \system{}’s feedback as moderately helpful (M = 3.00), and pointed out that some of the feedback could be improved. These concerns may reflect the limitations of current prompt-based approaches for building \system{}. Recent work shows that prompt-based LLM-generated texts often exhibit persistent, model-agnostic weaknesses such as clichés in writing quality \cite{chakrabarty_can_2025,chakrabarty_ai-slop_2025}, suggesting purely prompt-based approaches may inherently struggle with consistency. Future work could explore more robust techniques, such as finetuning on curated writing center dialogue, to better reflect pedagogical practices. 

Third, we reviewed \system{} as a complete system and did not perform ablations. This limits our ability to strictly attribute perceived benefits to specific design components. Future research can isolate how elements such as the balance of feedback types, prompts, and interaction flow independently shape users’ perceptions and experiences with systems.

Lastly, we acknowledge that writing center pedagogy represents just one of many possible pedagogical or composition frameworks for informing the design of writing support systems, alongside others such as genre-based or writing program administration approaches. We encourage future research to explore alternative educationally grounded approaches, particularly those from different cultural contexts, that may offer valuable alternatives or additions to the principles we propose.
% DONE final checking 

\section{Conclusion}

In this work, we demonstrate that writing center pedagogy provides a valuable theoretical foundation for designing AI writing support systems that address current concerns about plagiarism, reduced writer agency, and undermining critical thinking. Through interviews with writing tutors, we translated pedagogical principles into seven actionable design guidelines and implemented them in \system{}, an AI writing support prototype that provides non-directive feedback. Expert review with 30 domain experts revealed that \system{} was consistently perceived as preferred over other generative AI tools---especially among AI-skeptical educators. \system{} also aligns closely with writing center principles and shows potential complementary integration contexts that enhance rather than replace existing writing support. 
Building on these findings, we discuss design implications, including designing non-directive systems that earn trust among AI-skeptical educators, designing AI tools to integrate within the human writing ecosystem, integrating specific praises into feedback, and designing for adaptive granularity in feedback for future works.

\begin{acks}
The authors would like to thank Shangyun Wu for the teaser figure illustrations, the participants for their time and contributions, and the reviewers for their constructive feedback. The authors would also like to thank Dr. Carolyn Wisniewski, Dr. Kristina Aikens, and Ti-Chung Cheng for their thoughtful feedback on the project. 
\end{acks}

\bibliographystyle{ACM-Reference-Format}
\bibliography{ref}
\newpage
\appendix
%TC:ignore

\section{Interview Questions}
\label{sec:interview_guide}
\subsection*{Opening (5-8 min)}
\begin{itemize}
    \item Welcome and introduction.
    \item Brief overview of the research project and purpose of the interview.
    \item Present the consent form; assure confidentiality and explain that the interview will be recorded for research purposes; give time for the participants to ask questions about the consent form.
    \item Obtain consent.
\end{itemize}

\subsection*{Body (50 min)}

\subsubsection*{Tutor Background (5 min)}
\begin{itemize}
    \item Can you tell me about your experience as a writing tutor?
    \item How long? Where? Which grade level?
    \item What kinds of articles and students do you mostly work with? At what stages of writing?
\end{itemize}

\subsubsection*{Tutoring Approaches and Strategies (15 min)}
\begin{itemize}
    \item \textbf{[Grounded to a scenario the tutor described earlier]} We want to focus on the editing stage, where students come in and present a draft. What kinds of strategies do you usually use?
    \item Can you give examples of non-directive tutoring strategies, such as scaffolding, that you use for advising on a draft? How effective do you find these? Do you think students find these approaches useful?
    \item How do you balance offering guidance while ensuring that students retain ownership of their writing?
    \item What kinds of questions or prompts do you find most effective for helping students think critically about their writing?
    \item If you find out that a student might be using a language model to write their script, what guidance do you think is most necessary to give them? 
\end{itemize}

\subsubsection*{AI \& Writing (24 min)}
\begin{itemize}
    \item Do you use AI during your sessions? If so, how?
    \item How do you feel about students’ writing after ChatGPT and other large language models gained tremendous popularity? What are your opinions on these tools from a writing tutor's perspective?
    \item How might an AI writing support tool complement the work done in writing centers? Are you using any AI tools right now during your sessions?
    \item How could AI augment or help before, during, or after tutoring sessions with you?
    \item What opportunities do you see for expanding access to writing support through AI?
    \item How could an AI tool potentially address common issues you encounter in tutoring sessions?
\end{itemize}

\subsection*{Closing (6 min)}
\begin{itemize}
    \item Is there anything else you’d like to share about your experience as a tutor or your thoughts on integrating writing center strategies into AI systems?
    \item Do you have any concerns or suggestions for the direction of this research project?
    \item Based on your experience, what advice would you give to developers creating an AI writing support tool?
\end{itemize}

\subsection*{Conclusion (2 min)}
\begin{itemize}
    \item Thank the tutor for their time and insights.
    \item Explain the next steps in the research process, restating how the interview data will be used.
\end{itemize}

\section{System \& Prompts}
\label{sec:appendix}

The prompting strategy for areas of improvement follows the following 4-stage pipeline:

\begin{enumerate}
    \item Topic Identification (Fig.\ref{fig:Prompt_hoc}): Identifies high-order concerns (topics of writing concerns) based on the writer's selected goals (\textbf{G7}); 
    \item Sentence-Level Analysis (Fig.\ref{fig:Prompt_sentence}): Maps identified topics to specific sentences, focusing on the most significant issues (limited to top 5);
    \item Feedback Type Selection (Fig.\ref{fig:Prompt_feedback_type}): Determines the most appropriate feedback approach (e.g., reader-perspective feedback) for each identified sentence;
    \item Final Feedback Creation (Fig.\ref{fig:Prompt_final_feedback}): Generates concise (under 700 characters) feedback using the feedback type paired with open-ended questions to promote writer engagement.
\end{enumerate}

\begin{figure*}[ht]
    \begin{framed}
    \begin{flushleft} % Forces everything inside to the left
        \small \ttfamily

Given the following assignment details (with writer's goals), identify up to 4 major high-order concerns that are the most urgent ones to fix from. However, please also make sure to only identify the topics that are relevant to the goals of the assignment. High-order concerns are:
- Thesis/Argument: Whether the main argument is clear and well-structured.
- Organization: The logical flow and structure of ideas.
- Development: Whether evidence, examples, and reasoning support arguments.
- Audience and Purpose: How well the writing communicates its purpose to the intended audience.

This is the goals students want to achieve for this essay:
\{assignment\_goals\}

Return your response in JSON form, be specific in the reason:

\begin{verbatim}
{
    "HOCs": [
        {"Issue": ""},
        {"Issue": ""}
    ]
}
\end{verbatim}
\end{flushleft}
    \end{framed}
    \caption{Prompt for Topic Identification}
    \Description{Prompt for topic identification, including a definition of topic categories and a return format in JSON. Here's the full prompt: Given the following assignment details (with writer"s goals), identify up to 4 major high-order concerns that are the most urgent ones to fix from. However, please also make sure to only identify the topics that are relevant to the goals of the assignment. High-order concerns are: - Thesis/Argument: Whether the main argument is clear and well-structured. - Organization: The logical flow and structure of ideas. - Development: Whether evidence, examples, and reasoning support arguments. - Audience and Purpose: How well the writing communicates its purpose to the intended audience. This is the goals students want to achieve for this essay: {assignment_goals} Return your response in JSON form, be specific in the reason:{"HOCs": [{"Issue": ""}, {"Issue": ""}]}.}
    \label{fig:Prompt_hoc}
\end{figure*}

\begin{figure*}[ht]
    \centering
    \begin{framed}
    \begin{flushleft} % Forces everything inside to the left
        \small \ttfamily

The essays want to focus on the following issues due to the given reasons:
\{topic\_results\}

Please identify the issues in sentence level that are related to the issues and reasons.
If you believe that the issue requires inserting a new sentence, highlight the previous sentence of the new sentences to be inserted.

Here is the writing that needs to be improved:
\{essay\}

Return your response in JSON form, only return the top 5 most significant sentences' issues:

\begin{verbatim}
{
    "Sentences": [
        {
            "Sentence": "",
            "HOC": "",
            "Reason": ""
        },
        ...
    ]
}
\end{verbatim}
    \end{flushleft}
    \end{framed}
    \caption{Prompt for Sentence-Level Issues}
    \Description{Prompt for sentence-level issues, passing the topics identified earlier and the essay to identify the sentences with the issues, with a return format in JSON. Here's the full prompt: The essays want to focus on the following issues due to the given reasons: {topic_results}. Please identify the issues in sentence level that are related to the issues and reasons. If you believe that the issue requires inserting a new sentence, highlight the previous sentence of the new sentences to be inserted. Here is the writing that needs to be improved: {essay}. Return your response in JSON form, only return the top 5 most significant sentences" issues: { "Sentences": [ { "Sentence": "", "HOC": "", "Reason": "" }, ... ] }.}
    \label{fig:Prompt_sentence}
\end{figure*}

\begin{figure*}[ht]
    \centering
    \begin{framed}
    \begin{flushleft} % Forces everything inside to the left
        \small \ttfamily

We want to give two types of feedback: Reader-Perspective Feedback and feedback involving giving examples or analogies.
Giving reader-perspective feedback involves providing feedback on how the writing is perceived by the reader.
Giving examples or analogies involves providing examples or analogies to help the writer understand how to improve their writing.

Here are a list of problematic sentences that need feedback, with reasons and issue categories:
\{sentence\_results\}

Here is the argumentative essay:
\{essay\}

Please decide which type of feedback is more appropriate for each sentence and provide feedback accordingly.

Return your response in JSON form:

\begin{verbatim}
{
    "Feedback_type": [
        {
            "Sentence": "",
            "HOC": "",
            "Reason": "",
            "FeedbackType": ""
        },
        ...
    ]
}
\end{verbatim}

    \end{flushleft}
    \end{framed}
    \caption{Prompt for Feedback Type Selection}
    \Description{Prompt for feedback type selection, passing the results from the previous sentence-level issues, returning the feedback type to be given for the sentence and issue, with a return format in JSON. Here's the full prompt: We want to give two types of feedback: Reader-Perspective Feedback and feedback involving giving examples or analogies. Giving reader-perspective feedback involves providing feedback on how the writing is perceived by the reader. Giving examples or analogies involves providing examples or analogies to help the writer understand how to improve their writing. Here are a list of problematic sentences that need feedback, with reasons and issue categories: {sentence_results} Here is the argumentative essay: {essay} Please decide which type of feedback is more appropriate for each sentence and provide feedback accordingly. Return your response in JSON form: { "Feedback_type": [ { "Sentence": "", "HOC": "", "Reason": "", "FeedbackType": "" }, ... ] }.}
    \label{fig:Prompt_feedback_type}
\end{figure*}

\begin{figure*}[ht]
    \centering
    \begin{framed}
    \begin{flushleft} % Forces everything inside to the left
        \small \ttfamily

Please provide an empathetic and encouraging feedback for the following sentences. 
Please use the specific feedback type in the list:
\{type\_results\} 

Here is the complete argumentative essay:
\{essay\}

Keep in mind these are user's goals and assignment details:
\{assignment\_details\}

Return your response in JSON form, never give users anything to copy and paste directly into their essay:

Make sure each feedback sentence ends with an open-ended question to promote engagement, and keep the feedback under 700 characters. Keep HOCs into one or two word. 

\begin{verbatim}
{
    "Feedback": [
        {
            "Sentence": "",
            "HOC": "",
            "Reason": "",
            "FeedbackType": "",
            "Feedback": "",
        },
        ...
    ]
}
\end{verbatim}

    \end{flushleft}
    \end{framed}
    \caption{Prompt for Final Feedback Generation}
    \Description{Prompt for final feedback generation, asking for emphatic and encouraging feedback on specific sentences using the feedback types from {type_results}, applied to the full essay {essay} and assignment details {assignment_details}. The prompt instructs not to provide text that can be copied directly, to end each feedback with an open-ended question, and returned in JSON. Full prompt here: Please provide an emphatic and encouraging feedback for the following sentences. Please use the specific feedback type in the list: {type_results} Here is the complete argumentative essay: {essay} Keep in mind these are user"s goals and assignment details: {assignment_details} Return your response in JSON form, never give users anything to copy and paste directly into their essay. Make sure each feedback sentence ends with an open-ended question to promote engagement, and keep the feedback under 700 characters. Keep HOCs into one or two word. { "Feedback": [ { "Sentence": "", "HOC": "", "Reason": "", "FeedbackType": "", "Feedback": "" }, ... ] }.}
    \label{fig:Prompt_final_feedback}
\end{figure*}

\begin{figure*}[ht]
    \centering
    \begin{framed}
    \begin{flushleft} % Forces everything inside to the left
        \small \ttfamily

Please identify the sentences that the writer did well and provide encouraging feedback for them.

Here is the complete argumentative essay:
\{essay\}

The category should always include a praise word like "Good", "Excellent" plus the specific aspect.

Return your response in JSON form only for the top 3 most significant sentences, keep the feedback under 400 characters, be concise yet constructive.

\begin{verbatim}
{
    "Encouragement": [
        {
            "Sentence": "",
            "Feedback": "",
            "Category": ""
        },
        ...
    ]
}
\end{verbatim}

    \end{flushleft}
    \end{framed}
    \caption{Prompt for Encouragement Feedback}
    \Description{Prompt for encouragement feedback, asking to identify three places writers did good and generate a detailed praise, return in JSON format. Full prompt here: Please identify the sentences that the writer did well and provide encouraging feedback for them. Here is the complete argumentative essay: {essay} The category should always include a praise word like "Good", "Excellent" plus the specific aspect. Return your response in JSON form only for the top 3 most significant sentences, keep the feedback under 400 characters, be concise yet constructive. { "Encouragement": [ { "Sentence": "", "Feedback": "", "Category": "" }, ... ] }.}
    \label{fig:Prompt_encouragement}
\end{figure*}

\begin{figure*}[!htbp]
    \centering
    \begin{framed}
    \begin{flushleft} % Forces everything inside to the left
        \small \ttfamily
Here is the assignment prompt and requirements '\{assignment\_prompt\}', please analyze what the expected goals are for the writing to fit the prompt or grading rubrics. Now, the user also have their expectations for the editing service, which are: \{edit\_expectations\}. Given the information about the writing prompt, please provide the top 4 goals that the writer should aim for in their writing. Goal 5 should be a goal aim to satisfy the instructor/grader's expectations for the assignment, who is described as \{reader\}. Be specific in your goals, refrain from broad goals. Return ONLY a JSON object with the following structure:

\begin{verbatim}
{
  "goals": [
    "Goal 1",
    "Goal 2",
    "Goal 3",
    "Goal 4",
    "Goal 5"
  ]
}
\end{verbatim}

Replace the placeholder goals with the actual goals. Do not include any extra text.

    \end{flushleft}
    \end{framed}
    \caption{Prompt for Generating Goals}
    \Description{Prompt for generating goals in the goal-setting stage, taking the assignment details, edit expectations, and audience user inputted in the first page to generate five detailed goals, return in JSON format. Full prompt here: "Here is the assignment prompt and requirements {assignment_prompt}, please analyze what the expected goals are for the writing to fit the prompt or grading rubrics. Now, the user also has their expectations for the editing service, which are: {edit_expectations}. Given the information about the writing prompt, please provide the top 4 goals that the writer should aim for in their writing. Goal 5 should be a goal to satisfy the instructor/grader's expectations for the assignment, who is described as {reader}. Be specific in your goals, refrain from broad goals. Return ONLY a JSON object with the following structure: { "goals": [ "Goal 1", "Goal 2", "Goal 3", "Goal 4", "Goal 5" ] }".}

    \label{fig:Prompt_writinggoals}
\end{figure*}

\begin{figure*}[t]
    \centering
    \includegraphics[width=0.9\textwidth]{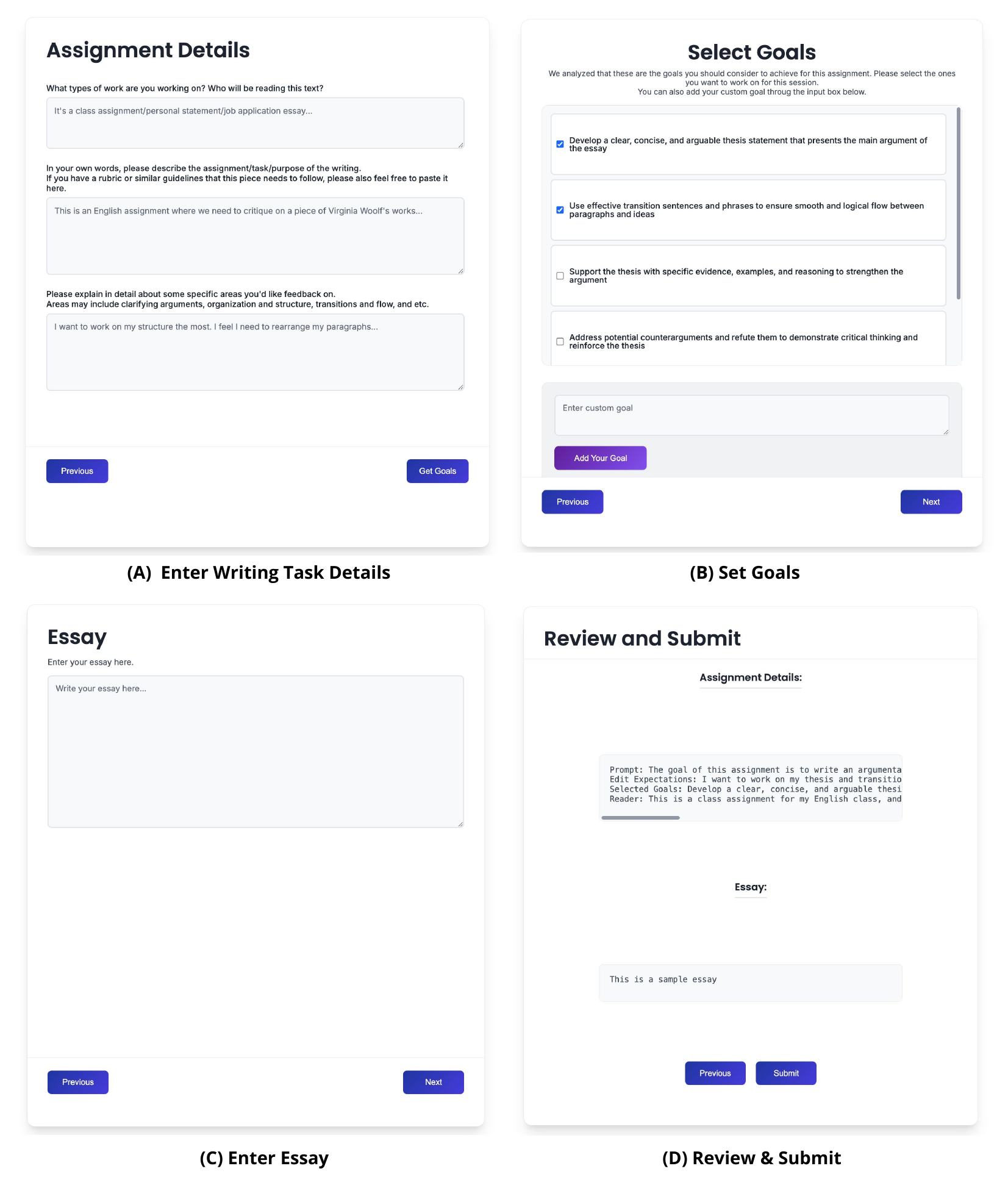}
    \caption{Screenshots for the Goal Setting Stage}
    \Description{Screenshots for the goal setting stage, with four subfigures showing (A) entering writing task details, with three questions asking about reader audience, assignment prompt, and edit expectations; (B) Select goals, showing five generated goals with a text box to input customized goals; (C) Entering the essay, including a text box for copy-paste; and (D) Review and submit, including one box to review the entered details and selected goals, and another box showing the essay.}
    \label{fig:interface-goal-real}
\end{figure*}

\newpage
\clearpage

\section{Internal Audit}
\label{sec:audit_questions}
\textbf{Goal Evaluation Definitions:}

\begin{itemize}
    \item \textbf{Relevant}: A goal is relevant if it clearly aligns with either the assignment prompt or the student’s stated edit expectations.
    \item \textbf{Specific}: A goal is specific if it mentions explicit concepts or strategies directly traceable to the prompt or edit expectations. If it is too vague or broad, it is marked “Not specific.”
    \item \textbf{Goal 5 Tailoring}: Goal 5 is tailored if it reflects awareness of the potential reader’s role (e.g., instructor, grader, or HR) and tries to match their likely priorities (e.g., clarity, argument depth, detail). If it’s a Goal 5, it will display as “to be filled” in the audit sheet. We have in total four Goal 5 evaluated. 
\end{itemize}

\textbf{Topic Evaluation Definitions:}

\begin{itemize}
    \item \textbf{Topic Fit}: Topics should fall into one of the following: Thesis/Argument, Organization, Development, Audience and Purpose.
    \item \textbf{Topic Alignment}: The topic should directly address a challenge the goals aim to improve. No alignment means it introduces a new issue.
\end{itemize}

\textbf{Critique Evaluation Definitions:}

\begin{itemize}
    \item \textbf{Appropriate}: Feedback should correctly identify and stay focused on the issue.
    \item \textbf{Non-Directive}: Feedback should encourage reflection rather than prescribe exact fixes.
    \item \textbf{Feedback Types}:
    \begin{itemize}
        \item \textit{Reader-Perspective}: Offers feedback on how the writing is perceived.
        \item \textit{Examples or Analogies}: Helps illustrate revision strategies.
    \end{itemize}
\end{itemize}

\textbf{Praise Evaluation Definitions:}
\begin{itemize}
\item \textbf{Appropriate:} Focuses on real strength.
\item \textbf{Specific:} References detailed content.
\end{itemize}

\subsection{Comparison to Baseline}
\label{sec:baseline}

We compared our prompting approach (e.g., 4-stage critique generation and separate praise generation) to a single prompt (Fig.~\ref{fig:prompt_praise_critique}). We compared it with a total of 150 critiques and 90 praises from 10 different essays as presented as samples used in the user study. We present our findings in Table~\ref{tab:final_audit}.

\begin{table*}[ht]
\scriptsize
\centering
\resizebox{\linewidth}{!}{
\begin{tabular}{@{}p{1.9cm}p{1cm}p{3.5cm}p{3.5cm}p{0.8cm}p{0.8cm}p{1.5cm}@{}}
\toprule
\textbf{Type} & \textbf{ID} & \textbf{Details} & \textbf{Generated Goal} & \textbf{Relevant?} & \textbf{Specific?} & \textbf{G5 Tailored?} \\
\midrule
Argumentative Essay & Essay 1 & Class assignment analyzing \textit{The Age of Innocence}. Prompt asks for contrast between “spaces of dreams” and societal norms. & Develop a clear and arguable thesis that addresses how Wharton contrasts these elements and what the Parisian setting reveals. & No & No & Not applicable \\
\bottomrule
\end{tabular}
}
\caption{Audit Example – Goal Generation}
\Description{One row showing an example for goal generation audit; information includes essay types, assignment details, and generated goals; evaluated on relevance, specificity, and alignment of Goal 5.}
\end{table*}

\begin{table*}[ht]
\scriptsize
\centering
\resizebox{\linewidth}{!}{
\begin{tabular}{@{}p{1.5cm}p{1cm}p{2.2cm}p{3cm}p{2.5cm}p{1.2cm}p{1.2cm}@{}}
\toprule
\textbf{Type} & \textbf{ID} & \textbf{Goal Summary} & \textbf{Generated Topical Concerns} & \textbf{Category} & \textbf{Fit?} & \textbf{Aligned?} \\
\midrule
Argumentative Essay & Essay 1 & Goal about thesis clarity & Lacks a concise thesis that addresses both dreams and Parisian symbolism. & Thesis/Argument & Yes & Yes \\
\bottomrule
\end{tabular}
}
\caption{Audit Example – Topic Mapping}
\Description{One row showing an example for topic mapping audit; information includes essay types, selected goal, generated writing concerns based on the goals, and the concern topic category; evaluated on whether the concern topic fits into predefined categories and on alignment between the concern and the goal.}
\end{table*}

\begin{table*}[ht]
\centering
\footnotesize
\begin{tabular}{@{}p{1.5cm}p{1.8cm}p{3cm}p{4.5cm}p{0.6cm}p{0.6cm}p{0.6cm}p{0.6cm}p{0.6cm}@{}}
\toprule
\textbf{Type} & \textbf{Feedback Type} & \textbf{Sentence for Feedback} & \textbf{Generated Response} & \textbf{A?} & \textbf{I?} & \textbf{U?} & \textbf{ND?} & \textbf{AL?} \\
\midrule
Argumentative Essay & Reader-Perspective & "In Edith Warton's The Age of Innocence, dreams of self-fulfillment lie in contrast to the context in which the characters exist—the time, the place, the society." & Topic: Thesis/Argument; From a reader's perspective, this sentence feels somewhat vague and does not clearly set up the specific argument the essay will make. Consider refining it to clearly state your main thesis that directly addresses both the contrast of the dream spaces versus Old New York society and how Paris symbolizes the hope or impossibility in Newland and Ellen's romance. This clarity will help readers immediately grasp your essay's focus. & Yes & No & Yes & Yes & Yes \\
\bottomrule
\end{tabular}
\caption{Audit Example - Critique Feedback Evaluation. A=Appropriate, I=Inaccurate, U=Usable Text, ND=Non-Directive, AT=Aligned Type}
\Description{One row showing an example for critique generation audit; information includes essay type, feedback type, sentence receiving the critique, and the generated critique; evaluated on appropriateness, accuracy, usability, non-directiveness, and alignment.}
\end{table*}

\begin{table*}[ht]
\centering
\footnotesize
\begin{tabular}{@{}p{1.4cm}p{3cm}p{3.5cm}p{0.8cm}p{0.8cm}@{}}
\toprule
\textbf{Type} & \textbf{Sentence for Feedback} & \textbf{Praise Feedback} & \textbf{Approp.?} & \textbf{Specific?} \\
\midrule
Argumentative Essay & “He wants a space devoid of context, not only without expectations and conventions to follow, but entirely without categories to qualify love.” & Excellent sentence that effectively unpacks complex ideas about societal constraints and personal desire with clarity and depth & Yes & Yes \\
\bottomrule
\end{tabular}
\caption{Audit Example – Praise Feedback}
\Description{One row showing an example for praise generation audit; information includes essay type, sentence receiving the praise, and the generated praise; evaluated on appropriateness and specificity.}
\end{table*}

\begin{figure*}[ht]
    \centering
    \begin{framed}
    \begin{flushleft} % Forces everything inside to the left
        \small \ttfamily
Please provide feedback on the following essay.

Below are the assignment details and the student's stated goals:
\{assignment\_details\}

Please review the essay and provide:
1. Three specific sentences that deserve praise (Encouragement).
2. Five specific sentences that need improvement (Critique), using non-directive and constructive feedback.

Here is the essay:
\{essay\}

Return your response in JSON form:

\begin{verbatim}
{
    "Praise": [
        { "Sentence": "...", "Feedback": "..." },
        ...
    ],
    "Critiques": [
        { "Sentence": "...", "Feedback": "..." },
        ...
    ]
}
\end{verbatim}

    \end{flushleft}
    \end{framed}
    \caption{Prompt for Baseline Praise and Critique Generation}
    \Description{Prompt for generating baseline praise and critique, taking the assignment details and the essay text to generate structured feedback. The prompt requests three specific sentences for praise and five for critique using non-directive constructive feedback, returned in JSON format. Full prompt here: Please provide feedback on the following essay. Below are the assignment details and the student’s stated goals: {assignment_details} Please review the essay and provide: 1. Three specific sentences that deserve praise (Encouragement). 2. Five specific sentences that need improvement (Critique), using non-directive and constructive feedback. Here is the essay: {essay} Return your response in JSON form: {"Praise": [{ "Sentence": "...", "Feedback": "..." }, ... ], "Critiques": [{ "Sentence": "...", "Feedback": "..." }, ... ] }}
    \label{fig:prompt_praise_critique}
\end{figure*}

\begin{table*}[t]
\centering
\small
\begin{tabular}{l l r r r r c}
\toprule
\textbf{Type} & \textbf{Metric} & \textbf{Writor} & \textbf{Baseline} & \textbf{Diff \%} & \textbf{P-Value} & \textbf{Sig.} \\
\midrule
Critique & Length      & 36.67 & 24.96 & +46.9\% & 0.0000 & ** \\
Critique & Specificity & 10.32 & 5.92  & +74.3\% & 0.0000 & ** \\
Critique & Sentiment   & 0.25  & 0.21  & +14.8\% & 0.1599 &  \\
Praise   & Length      & 21.14 & 19.89 &  +6.3\% & 0.0414 & * \\
Praise   & Specificity & 6.88  & 5.72  & +20.2\% & 0.0013 & ** \\
Praise   & Sentiment   & 0.42  & 0.28  & +51.9\% & 0.0001 & ** \\
\bottomrule
\end{tabular}
\caption{Comparison between Writor's feedback and a single-prompt baseline (N=10 essays, 3 runs each) using paired t-test. Specificity measures the average number of noun chunks (concepts) per feedback item. $^{**}p<0.01, ^{*}p<0.05$. }
\label{tab:final_audit}
\Description{Comparison of Writor’s feedback versus a baseline prompt across praises and critiques. Critique metrics showed significant increases for Writor in specificity and length, but no significant difference in Sentiment. Praise metrics showed significant increases for Writor across all categories: sentiment, specificity, and length.}
\end{table*}

\clearpage

\section{\system{} Evaluation Survey}
\label{sec:eval_survey}

The majority of survey questions used 5-point Likert-style scales, with labels adapted to match question intent. These included quality-oriented scales (Poor---Excellent), adoption likelihood (Definitely Not---Definitely Yes), agreement (Strongly Disagree---Strongly Agree), helpfulness (Not at All Helpful---Extremely Helpful), and comparative judgments (Much Worse---Much Better). For questions that needed educational background or system building background, we added a ``not applicable'' option. For example, we recognized that some AI researchers may not have experience with writing centers. Therefore, when asking participants to compare \system{} to writing centers, we included the ``not applicable'' option.

\subsection{Popup Survey}

\begin{enumerate}
    \item \textbf{Overall, how helpful was the feedback you received?} (Required)
    \begin{enumerate}
        \item Not at all helpful
        \item Slightly helpful
        \item Somewhat helpful
        \item Very helpful
        \item Extremely helpful
    \end{enumerate}

    \item \textbf{What impact do you think the system you interacted with, or systems like it, could have on writing education?} (Required)

    \item \textbf{What features of the system were most and least useful to you? Why?} (Required)

    \item \textbf{Do you mainly work in writing tutoring/teaching or AI writing? (Select all that apply)} (Required)
    \begin{enumerate}
        \item[$\square$] I teach writing
        \item[$\square$] I tutor
        \item[$\square$] I research or build AI writing systems
        \item[$\square$] None of the above
    \end{enumerate}

    \item \textbf{Please leave your email address (optional):}

\end{enumerate}

% Long Survey
\subsection{System's Feedback}

\begin{enumerate}
    \item \textbf{Accuracy}
    \begin{enumerate}
        \item Poor
        \item Fair
        \item Good
        \item Very good
        \item Excellent
        \item Not applicable
    \end{enumerate}

    \item \textbf{Relevance}
    \begin{enumerate}
        \item Poor
        \item Fair
        \item Good
        \item Very good
        \item Excellent
        \item Not applicable
    \end{enumerate}

    \item \textbf{Specificity (level of detail you wanted)}
    \begin{enumerate}
        \item Poor
        \item Fair
        \item Good
        \item Very good
        \item Excellent
        \item Not applicable
    \end{enumerate}

    \item \textbf{Clarity}
    \begin{enumerate}
        \item Poor
        \item Fair
        \item Good
        \item Very good
        \item Excellent
        \item Not applicable
    \end{enumerate}

    \item \textbf{Actionability}
    \begin{enumerate}
        \item Poor
        \item Fair
        \item Good
        \item Very good
        \item Excellent
        \item Not applicable
    \end{enumerate}

    \item \textbf{Balance (positive and critical feedback)}
    \begin{enumerate}
        \item Poor
        \item Fair
        \item Good
        \item Very good
        \item Excellent
        \item Not applicable
    \end{enumerate}

    \item \textbf{How would you describe the overall tone of the feedback?}
    \begin{enumerate}
        \item Very critical
        \item Somewhat critical
        \item Neutral
        \item Somewhat positive
        \item Very positive
    \end{enumerate}

    \item \textbf{Would you consider adopting this feedback approach in your own teaching or tutoring practice?}
    \begin{enumerate}
        \item Definitely not
        \item Probably not
        \item Unsure
        \item Probably yes
        \item Definitely yes
        \item Not applicable
    \end{enumerate}
\end{enumerate}

\textbf{Please answer the following questions. Feel free to skip any questions that are not relevant to you. Rate your agreement with the following statements:}

\begin{enumerate}
    \item \textbf{The feedback helped me boost my confidence in writing.}
    \begin{enumerate}
        \item Strongly disagree
        \item Somewhat disagree
        \item Neither agree nor disagree
        \item Somewhat agree
        \item Strongly agree
        \item Not applicable/Didn't See/Didn't Use
    \end{enumerate}

    \item \textbf{The feedback maintained the integrity of my ideas and voice.}
    \begin{enumerate}
        \item Strongly disagree
        \item Somewhat disagree
        \item Neither agree nor disagree
        \item Somewhat agree
        \item Strongly agree
        \item Not applicable/Didn't See/Didn't Use
    \end{enumerate}

    \item \textbf{The feedback helped me think more deeply about my writing.}
    \begin{enumerate}
        \item Strongly disagree
        \item Somewhat disagree
        \item Neither agree nor disagree
        \item Somewhat agree
        \item Strongly agree
        \item Not applicable/Didn't See/Didn't Use
    \end{enumerate}

    \item \textbf{The follow-up chat function for each feedback is helpful.}
    \begin{enumerate}
        \item Strongly disagree
        \item Somewhat disagree
        \item Neither agree nor disagree
        \item Somewhat agree
        \item Strongly agree
        \item Not applicable/Didn't See/Didn't Use
    \end{enumerate}

    \item \textbf{How helpful did you find this system's reader point of view ("As a reader, I noticed…")?}
    \begin{enumerate}
        \item Not at all helpful
        \item Slightly helpful
        \item Somewhat helpful
        \item Very helpful
        \item Extremely helpful
        \item Not applicable/Didn't See/Didn't Use
    \end{enumerate}

    \item \textbf{How useful was the function to highlight and receive feedback for customized sections?}
    \begin{enumerate}
        \item Not at all useful
        \item Slightly useful
        \item Somewhat useful
        \item Very useful
        \item Extremely useful
    \end{enumerate}
\end{enumerate}

\textbf{How would you compare the quality of the system's feedback to other forms of writing feedback you've received?}

\begin{enumerate}
    \item \textbf{Compared to instructor/professor feedback}
    \begin{enumerate}
        \item Much worse
        \item Worse
        \item About the same
        \item Better
        \item Much better
        \item Not applicable
    \end{enumerate}

    \item \textbf{Compared to peer feedback}
    \begin{enumerate}
        \item Much worse
        \item Worse
        \item About the same
        \item Better
        \item Much better
        \item Not applicable
    \end{enumerate}

    \item \textbf{Compared to writing center feedback}
    \begin{enumerate}
        \item Much worse
        \item Worse
        \item About the same
        \item Better
        \item Much better
        \item Not applicable
    \end{enumerate}

    \item \textbf{Compared to generative AI writing systems (ChatGPT, Claude, Gemini, DeepSeek, etc.)}
    \begin{enumerate}
        \item Much worse
        \item Worse
        \item About the same
        \item Better
        \item Much better
        \item Not applicable
    \end{enumerate}

    \item \textbf{Did you receive any feedback that you can directly copy to your text?}
    \begin{enumerate}
        \item Yes
        \item No
    \end{enumerate}
\end{enumerate}

\subsection{Interface and Designs for Writing Process}

\textbf{Please rate your agreement with the following statements:}

\begin{enumerate}
    \item \textbf{The system's feedback approach felt more like a conversation than a critique}
    \begin{enumerate}
        \item Strongly disagree
        \item Somewhat disagree
        \item Neither agree nor disagree
        \item Somewhat agree
        \item Strongly agree
    \end{enumerate}

    \item \textbf{I learned writing strategies that I can apply in the future}
    \begin{enumerate}
        \item Strongly disagree
        \item Somewhat disagree
        \item Neither agree nor disagree
        \item Somewhat agree
        \item Strongly agree
    \end{enumerate}

    \item \textbf{The system can support my development as a writer}
    \begin{enumerate}
        \item Strongly disagree
        \item Somewhat disagree
        \item Neither agree nor disagree
        \item Somewhat agree
        \item Strongly agree
    \end{enumerate}
\end{enumerate}

\subsection{Demographic Data}

\begin{enumerate}
    \item \textbf{What's your age range?}
    \begin{enumerate}
        \item 18-24
        \item 25-34
        \item 35-44
        \item 45-54
        \item 55-64
        \item 65 or above
    \end{enumerate}

    \item \textbf{What is your race/ethnicity?}
    \begin{enumerate}
        \item White
        \item Black or African American
        \item Asian
        \item Hispanic or Latino
        \item Native American or Alaska Native
        \item Native Hawaiian or Other Pacific Islander
        \item Other (Please specify): 
        \item Prefer not to say
    \end{enumerate}

    \item \textbf{What's your gender?}
    \begin{enumerate}
        \item Male
        \item Female
        \item Non-binary
        \item Prefer to self-identify: 
        \item Prefer not to say
    \end{enumerate}
\end{enumerate}

\subsection{Professional Background}

\begin{enumerate}
    \item \textbf{Years of experience (tutoring or teaching writing; researching or building AI writing systems):}
    \begin{enumerate}
        \item Less than 1 year
        \item 1 \quad \item 2 \quad \item 3 \quad \item 4 \quad \item 5
        \item 6 \quad \item 7 \quad \item 8 \quad \item 9 \quad \item 10
        \item 11 \quad \item 12 \quad \item 13 \quad \item 14 \quad \item 15
        \item 15+
    \end{enumerate}

    \item \textbf{Institution type (select all that apply):}
    \begin{enumerate}
        \item[$\square$] Public university
        \item[$\square$] Private university
        \item[$\square$] Industry
        \item[$\square$] Other (please specify): 
    \end{enumerate}

    \item \textbf{Your role (select all that apply):}
    \begin{enumerate}
        \item[$\square$] Writing center tutor
        \item[$\square$] Non-tenure track instructor
        \item[$\square$] Tenure-track/tenured faculty
        \item[$\square$] Research/Engineer roles in industry
        \item[$\square$] Other (please specify): 
    \end{enumerate}

    \item \textbf{How frequently do you use AI writing systems?}
    \begin{enumerate}
        \item Never
        \item Rarely (a few times a year)
        \item Occasionally (a few times a month)
        \item Regularly (a few times a week)
        \item Frequently (almost daily)
    \end{enumerate}

    \item \textbf{What is your attitude toward using AI for writing assistance?}
    \begin{enumerate}
        \item Very negative
        \item Somewhat negative
        \item Neutral
        \item Somewhat positive
        \item Very positive
    \end{enumerate}

    \item \textbf{How do your attitudes might differ between the system you just interacted with and other AI writing systems you have used?}

\end{enumerate}

\subsection{Follow-up Interview}

\begin{enumerate}
    \item \textbf{Would you be willing to enter the raffle of winning one of 5 \$25 Amazon Gift Card?} (Required)
    \begin{enumerate}
        \item Yes
        \item No
    \end{enumerate}

    \item \textbf{Please include your email if you would be interested in following up with us for a future interview.}
    \begin{enumerate}
        \item Yes
        \item No
    \end{enumerate}

    \item \textbf{If you answer yes to any of the questions above, please provide your email address:}

\end{enumerate}

\clearpage

\section{Additional Survey Results}
\label{sec:append_results}

\begin{table}[ht!]
\centering
\caption{Participants’ Professional Backgrounds}
\Description{Participants’ professional backgrounds breakdown; information includes specific role combinations (e.g., teaching plus tutoring or only AI research) and any role category. Most participants only taught writing (47\%), followed by teaching plus tutoring (20\%).}
\label{tab:professional-role-combinations}
\begin{tabular}{lcc}
\toprule
\textbf{Role Configuration} & \textbf{Count} & \textbf{Percentage} \\
\midrule
Writing Instructor (sole role) & 14 & 46.67\% \\
Writing Center Tutor (sole role) & 4 & 13.33\% \\
AI researcher (sole role) & 3 & 10.00\% \\
Instructor + Tutor & 6 & 20.00\% \\
Instructor + AI researcher & 2 & 6.67\% \\
Tutor + AI researcher & 0 & 0.00\% \\
All three roles & 1 & 3.33\% \\
\midrule
\textit{Any writing instructor experience} & 23 & 76.67\% \\
\textit{Any writing center tutoring experience} & 11 & 36.67\% \\
\textit{Any AI research experience} & 6 & 20.00\% \\
\bottomrule
\end{tabular}
\end{table}

\begin{figure*}[ht]
    \centering
    \includegraphics[width=0.8\linewidth]{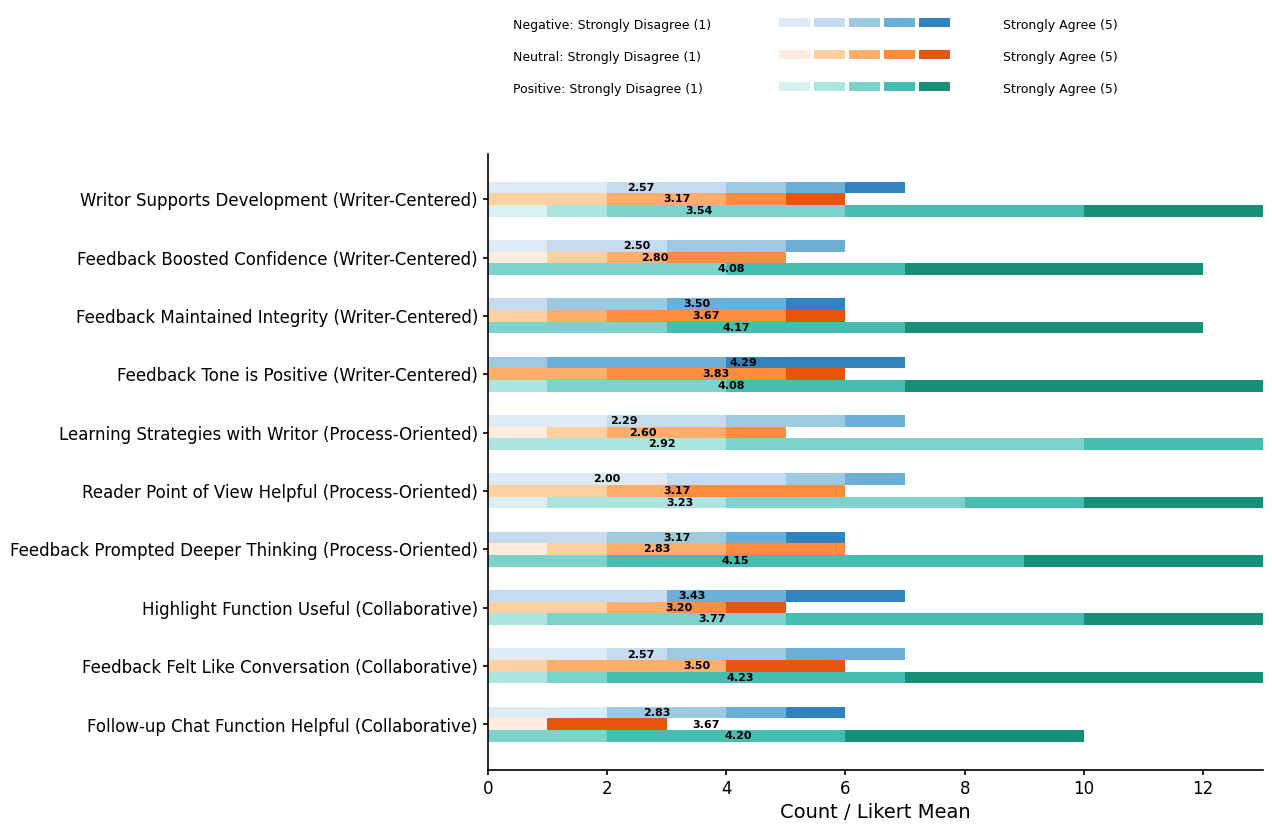}
    \caption{Participant ratings of \system{}'s alignment with writing center pedagogical principles separated by AI attitudes. The text overlaying on each bar indicates the mean rating score.}
    \label{fig:wc_attitude}
    \Description{Stacked horizontal bar chart of pedagogical alignment ratings grouped by AI attitude. Dimensions are categorized into Writer-Centered, Process-Oriented, and Collaborative. Positive-attitude participants generally rate highest across all categories. "Feedback tone is positive" received the highest ratings. }
\end{figure*}

\begin{figure*}[ht]
    \centering
    \includegraphics[width=0.8\linewidth]{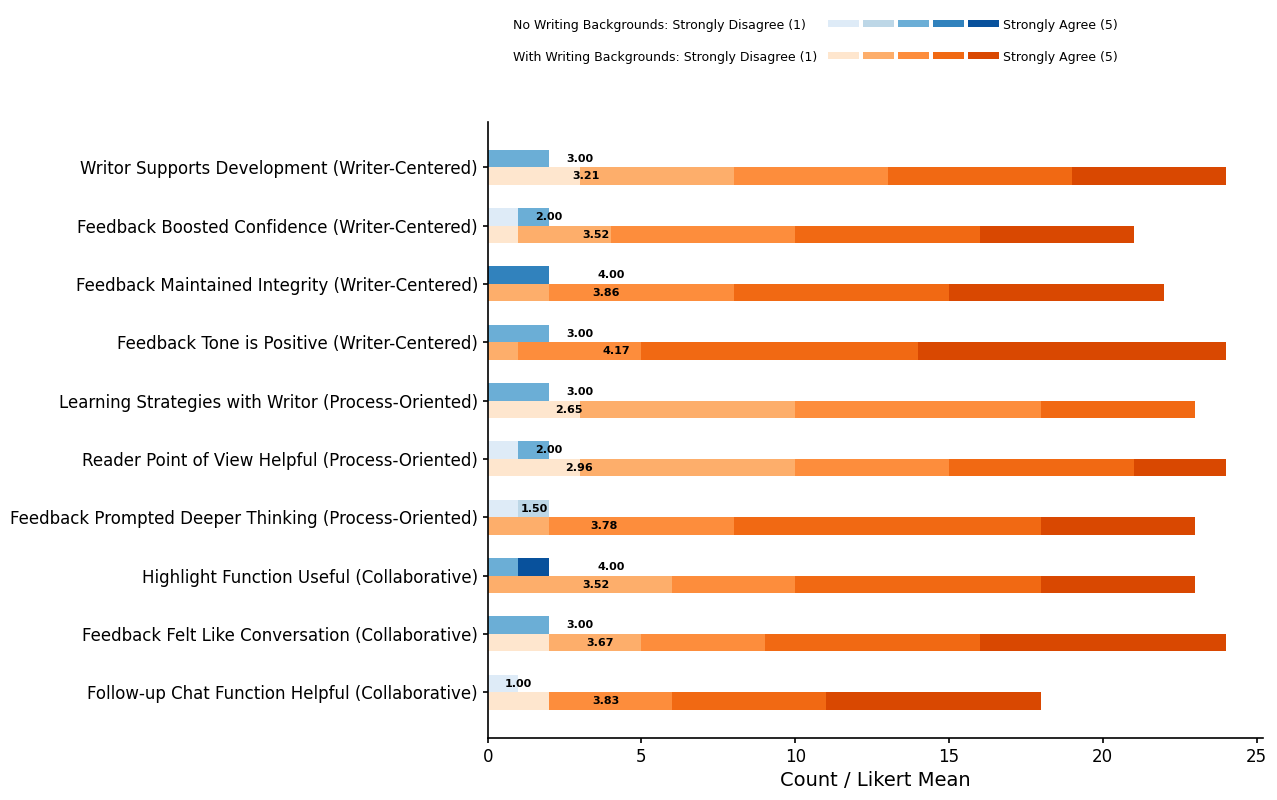}
    \caption{Participant ratings of \system{}'s alignment with writing center pedagogical principles separated by with or without writing instruction/tutoring backgrounds. ``No Writing Backgrounds'' group consists of participants who exclusively identified as AI researchers. The text overlaying on each bar indicates the mean rating score.}
    \label{fig:wc_role}
    \Description{Stacked horizontal bar chart of pedagogical alignment ratings grouped by participants with and without writing instruction backgrounds. Dimensions are categorized into Writer-Centered, Process-Oriented, and Collaborative. Participants with writing backgrounds rated ‘Feedback Tone is Positive’ (4.17) higher than those without (3.00). Conversely, participants without writing backgrounds gave higher ratings to ‘Feedback Maintained Integrity’ (4.00 vs 3.86). Nevertheless, these categories are all highly rated in their respective group.}
\end{figure*}

\begin{figure*}[ht]
    \centering
    \includegraphics[width=0.7\linewidth]{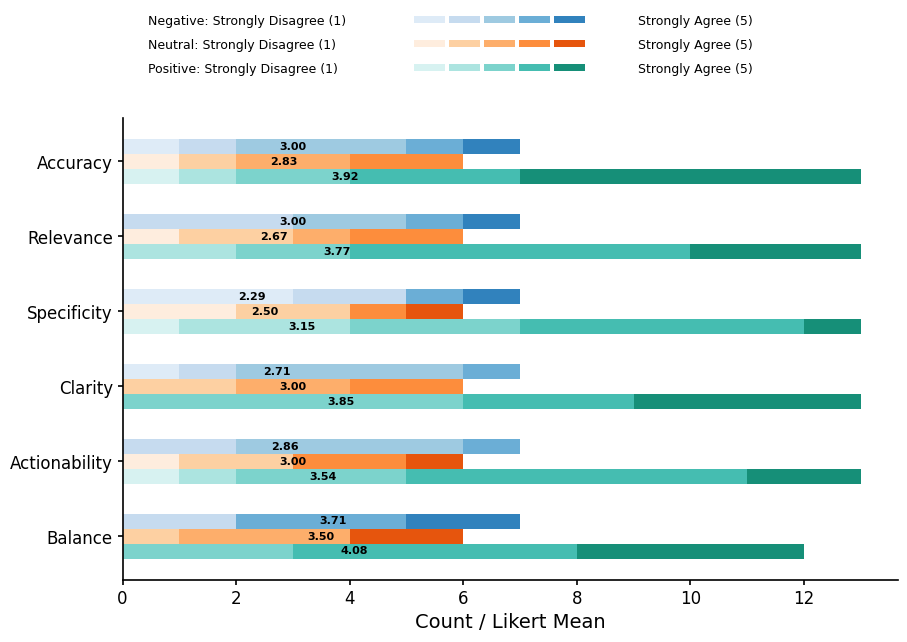}
    \caption{Participant ratings of \system{}'s feedback quality across six dimensions by AI attitudes. The text overlaying on each bar indicates the mean rating score.}
    \label{fig:quality_attitude}
    \Description{Stacked horizontal bar chart of feedback quality ratings across six dimensions (Specificity, Relevance, Clarity, Accuracy, Actionability, and Balance) grouped by AI attitude (Negative, Neutral, Positive). Positive-attitude participants consistently rate highest, followed by Neutral and Negative. "Balance" is the highest-rated metric across all groups, while ‘Specificity’ is the lowest.}
\end{figure*}

\begin{figure*}[ht]
    \centering
    \includegraphics[width=0.7\linewidth]{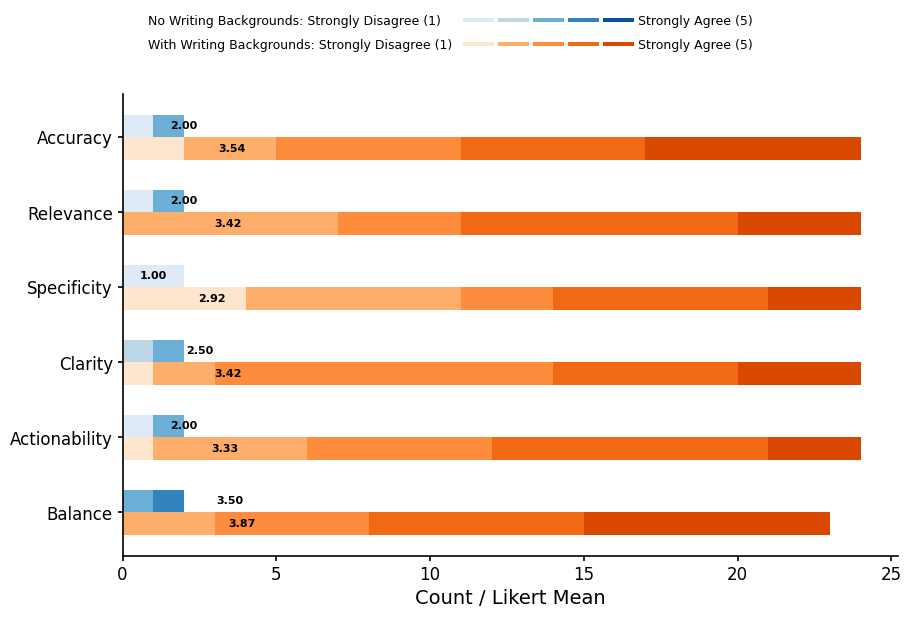}
    \caption{Participant ratings of \system{}'s feedback quality across six dimensions by with or without writing instruction/tutoring backgrounds. ``No Writing Backgrounds'' group consists of participants who exclusively identified as AI researchers. The text overlaying on each bar indicates the mean rating score.}
    \label{fig:quality_role}
    \Description{Stacked horizontal bar chart of feedback quality ratings across six dimensions (Specificity, Relevance, Clarity, Accuracy, Actionability, and Balance) grouped by writing instruction backgrounds. Participants with writing backgrounds consistently gave higher mean ratings across all dimensions compared to those without. "Balance" received the highest rating from both groups.}
\end{figure*}
%TC:endignore

\end{document}